\documentclass[aps,pra,twocolumn,showpacs,superscriptaddress,floatfix]{revtex4}
\usepackage{graphicx}
\usepackage{times}
\usepackage{nicefrac}
\usepackage{amsmath}
\usepackage{amsfonts}
\usepackage{amssymb}
\usepackage{amsthm}
\usepackage{epsf}
\usepackage{bm}
\usepackage{bbm}
\usepackage{longtable}

\usepackage{dcolumn}

\newcolumntype{.}{D{x}{}{-1}}

\newcommand{\vare}{\varepsilon}

\newcommand{\pr}{^{\prime}}

\newcommand{\hx}{\hat{\bfx}}

\newcommand{\bfr}{{\bf r}}
\newcommand{\bfx}{{\bf x}}

\newcommand{\balpha}{\bm{\alpha}}

\newcommand{\intinf}{\int^{\infty}_{-\infty}}
\newcommand{\lbr}{\langle}
\newcommand{\rbr}{\rangle}

\newcommand{\I}[4]{I_{#1\,#2\,#3\,#4}}
\newcommand{\DI}[4]{I^{\prime}_{#1\,#2\,#3\,#4}}
\newcommand{\U}[2]{U_{#1\,#2}}

\newcommand{\Za}{{Z \alpha}}
\newcommand{\im}{i}

\def\rms{<\!\!\!\,r^2\!\!\!>^{1/2}}

\begin{document}

\title{QED treatment of electron correlation in Li-like ions}

\author{V.~A.~Yerokhin}
 \affiliation{Center for Advanced Studies, St.~Petersburg State Polytechnical
University, Polytekhnicheskaya 29, St.~Petersburg 195251, Russia}
 \affiliation{Max--Planck Institut f\"ur Kernphysik, Saupfercheckweg 1, 69117
Heidelberg, Germany}

\author{A.~N.~Artemyev}
\affiliation{Department of Physics, St.~Petersburg State University, Oulianovskaya 1, Petrodvorets,
St.~Petersburg 198504, Russia}
\affiliation{Institut f\"ur Theoretische Physik, TU Dresden, Mommsenstrasse 13, D-01062 Dresden, Germany}

\author{V.~M.~Shabaev}
\affiliation{Department of Physics, St.~Petersburg State University, Oulianovskaya 1, Petrodvorets,
St.~Petersburg 198504, Russia}

\begin{abstract}

A systematic QED treatment of electron correlation is presented for ions along the lithium
isoelectronic sequence. We start with the zeroth-order approximation that accounts for a part of
the electron-electron interaction by a local model potential introduced into the Dirac equation.
The residual electron correlation is treated by perturbation theory. Rigorous QED evaluation is
presented for the first two terms of the perturbative expansion; the third-order contribution is
calculated within the many-body perturbation theory. We report accurate numerical results for the
electronic-structure part of the ionization potential for the $n=2$ states of all Li-like ions up
to uranium.

\end{abstract}


\maketitle
%%%%%%%%%%%%%%%%%%%%%%%%%%%%%%%%%%%%%%%%%%%%%%%%%%%%%%%%%%%%%%%%%%%%%%%
%
%%%%%%%%%%%%%%%%%%%%%%%%%%%%%%%%%%%%%%%%%%%%%%%%%%%%%%%%%%%%%%%%%%%%%%%

\section{Introduction}
\label{sec:intro}

Li-like ions are among the most fundamental many-electron systems. Excellent experimental accuracy
achieved for these ions and their relative simplicity make them
a very attractive object for an {\em ab initio} theoretical description. Numerous
measurements were performed for the Lamb shift for ions along the lithium 
isoelectronic sequence; some recent ones were presented in
Refs.~\cite{beiersdorfer:98,bosselmann:99,feili:00,madzunkov:02,brandau:04,kieslich:04,beiersdorfer:05}.
Large theoretical effort was invested during the last decade into rigorous calculations of QED
effects for the $2p_J$-$2s$ transition energies
\cite{blundell:93:a,yerokhin:99:sescr,artemyev:99,%
yerokhin:00:prl,yerokhin:01:2ph,sapirstein:01:lamb,andreev:01,%
artemyev:03,yerokhin:05:OS,yerokhin:06:prl}, which resulted in a substantial improvement of
the description of these transitions. However, the accuracy achieved in theoretical
investigations remained in most cases lower than that of the best 
experimental results. Its further improvement is important since a comparison of theoretical and
experimental results on a better level of accuracy will provide a test of QED effects up to second
order in the fine structure constant.

In the present investigation, we restrict ourself to an {\em ab initio} description of the
electronic-structure part of the energies of the $n=2$ states of Li-like ions. By the
electronic-structure part we understand the contributions that are induced by Feynman diagrams
involving only the electron-electron interaction (thus excluding the diagrams of the self-energy
and vacuum-polarization type) taken in the limit of the infinitely heavy nucleus. The
electronic-structure effects correspond to a well-defined and gauge-invariant part of the energy
and thus can be addressed separately. In order to obtain accurate theoretical predictions for the
energy levels, other effects should be added to the electronic-structure part, which comprise
the one-loop self-energy and vacuum-polarization (see, {\em e.g.}, the review \cite{mohr:98}), the
screening of one-loop self-energy and vacuum-polarization
\cite{yerokhin:99:sescr,artemyev:99,yerokhin:05:OS}, the two-loop QED effects
\cite{yerokhin:06:prl}, and the relativistic recoil effect \cite{artemyev:95:pra,artemyev:95:jpb}.

Within QED, a single electron-electron interaction is represented as an exchange of a virtual
photon and is given (in relativistic units $\hbar=c=1$) by the operator
\begin{equation}\label{eq1}
    I(\omega) = e^2\,\alpha_1^\mu\alpha_2^\nu\,D_{\mu\nu}(\omega,\bfx_{12})\,,
\end{equation}
where $D^{\mu\nu}$ is the photon propagator, whose expression in the Feynman gauge reads
\begin{equation}\label{eq2}
   D_{\mu\nu}(\omega,\bfx_{12}) = g_{\mu\nu}\, \frac{\exp \left( \im \sqrt{\omega^2+\im 0}\,x_{12}\right)}
          {4\pi x_{12}}\,,
\end{equation}
where $x_{12} = |\bfx_{12}| = |\bfx_1-\bfx_2|$ and $\alpha^{\mu} = (1,\balpha)$ are the Dirac
matrices. The electronic-structure part of the energy arises through exchanges of an arbitrary
number of photons.

The traditional approach to the treatment of the electron correlation (beyond the lowest-order
case of the one-photon exchange, which is rather simple) is based on a simplified form of the
interaction obtained within the Breit approximation. It is achieved by taking the
small-$\omega$ limit of the operator $I(\omega)$ while working in the Coulomb gauge (i.e.,
neglecting the retardation of the photon). We will indicate the Breit approximation employed for
the operator $I(\omega)$ simply by omitting the energy argument, $I(\omega)\to I$. The
corresponding expression consists of two parts, referred to as the Coulomb and the Breit
interaction,
\begin{eqnarray}\label{eq3}
&\displaystyle  I = I^{\rm Coul}+I^{\rm Breit}\,, \\
 &\displaystyle  I^{\rm Coul} = \frac{\alpha}{x_{12}}\,, \\
\label{eq3a}
 &\displaystyle  I^{\rm Breit}= -\frac{\alpha}{2\,x_{12}}\,
    \left[ \balpha_1\cdot\balpha_2 + \left( \balpha_1\cdot \hx_{12}\right)
             \left( \balpha_2\cdot \hx_{12}\right) \right]\,,
\end{eqnarray}
where $\hx = \bfx/|\bfx|$ and $\alpha = e^2/(4\pi)$ is the fine-structure constant.

The basic traditional methods for the relativistic structure calculations are the many-body
perturbation theory (MBPT) applied to the lithium isoelectronic sequence by Johnson {\em et al.}
\cite{johnson:88:b}, the multiconfigurational Dirac-Fock method applied by Indelicato and Desclaux
\cite{indelicato:90}, and the configuration-interaction method applied to Li-like ions by Chen {\em
et al.} \cite{chen:95}. All these methods treat the one-photon exchange correction exactly and the
higher-order electron correlation, within the Breit approximation only. They are thus incomplete to
the order $\alpha^2(\Za)^3 m$, which is the leading order for the two-electron QED effects. Rigorous
QED calculations of the two-photon exchange correction were performed for $n=2$ states of Li-like
ions in Refs.~\cite{yerokhin:00:prl,yerokhin:01:2ph,sapirstein:01:lamb,andreev:01,artemyev:03}. 
An exchange of
three and more more virtual photons can presently be treated within the Breit approximation only,
which leads to an incompleteness at the nominal order $\alpha^3(\Za)^2 m$. 

The three-photon exchange correction can be either calculated directly, as a quantum mechanical 
third-order perturbation correction with the same starting potential as in QED calculations, or
inferred from the total energies obtained in relativistic-structure calculations, by
subtracting the zeroth-, first-, and second-order perturbation terms evaluated separately. Care
should be taken in the latter case since the subtraction procedure involves large numerical 
cancellations.

Calculations of the three-photon exchange correction for Li-like ions were performed previously 
in Refs.~\cite{zherebtsov:00,andreev:01}. They were carried out on the Coulomb wave functions and the
results can be directly added to the existing values for the two-electron QED effects. Another
approach to this problem is advocated by Sapirstein and Cheng \cite{sapirstein:01:lamb} and
consists in starting the perturbative expansion with a local model potential, 
which incorporates a part of the electron-electron interaction effects. By a proper
choice of the model potential, one can significantly accelerate the convergence of the perturbative
expansion. In Ref.~\cite{sapirstein:01:lamb}, this approach was applied to the case of Li-like
bismuth and the results are in good agreement with those obtained with the Coulomb wave functions
\cite{yerokhin:00:prl,yerokhin:01:2ph}. Advantages of this approach should become more
evident for the smaller values of $Z$, where the convergence of the perturbative expansion becomes
slower.

In the present investigation, we adopt the method of Ref.~\cite{sapirstein:01:lamb} and apply it to
the study of the electronic-structure effects for ions along the isoelectronic sequence of lithium.
The approach can be regarded as a successor of the MBPT treatment applied to Li-like ions by
Johnson {\em et al.} \cite{johnson:88:b}; the difference is that we perform a rigorous QED
evaluation of the second-order correction and employ a different potential for the zeroth-order
approximation.

The paper is organized as follows. The choice of different model potentials to be used for the
zeroth-order approximation are discussed in the next section. In Sec.~\ref{sec:mbpt}, the 
perturbative expansion for the electron-correlation effects is constructed within the
MBPT approximation. We present results of numerical calculations for different model
potentials in the case of lithium and compare the convergence of the resulting expansions. The
rigorous QED calculation of the electron correlation through the second order of perturbation
theory is presented in Sec.~\ref{sec:qed}. We combine the QED values for the two-photon exchange
correction with the MBPT results from the previous section to obtain the electronic-structure part
of the ionization potential for the $n=2$ states of atoms along the lithium isoelectronic
sequence. In the last section, we collect all theoretical contributions available for the
$2p_{1/2}$-$2s$ and $2p_{3/2}$-$2s$ transition energies of several Li-like ions, compare them with
experimental results, and discuss perspectives for further improvement of theoretical
predictions.

%%%%%%%%%%%%%%%%%%%%%%%%%%%%%%%%%%%%%%%%%%%%%%%%%%%%%%%%%%%%%%%%%%%%%%%
%
%%%%%%%%%%%%%%%%%%%%%%%%%%%%%%%%%%%%%%%%%%%%%%%%%%%%%%%%%%%%%%%%%%%%%%%

\section{Choice of potential}
\label{sec:pot}

We consider here three local model potentials that are supposed to account for a part of the
interaction between the valence electron and the closed core. The simplest choice is the
core-Hartree (CH) potential defined as
\begin{equation}\label{eq4}
    V_{\rm CH}(r) = V_{\rm nuc}(r) + \alpha \int_0^{\infty} dr\pr \frac1{r_>}\, \rho_c^{s.c.}(r\pr)\,,
\end{equation}
where $V_{\rm nuc}(r)$ denotes the nuclear potential ({\em i.e.}, the Coulomb potential induced by
an extended nuclear-charge distribution), $r_> = \max(r,r\pr)$, $\rho_c$ is the density of the core
electrons,
\begin{equation}\label{eq5}
    \rho_c(r) = \sum_{n_c} (2j_c+1) \left[ G_c^2(r)+F_c^2(r)\right]\,,
\end{equation}
the superscript ``$s.c.$" indicates that the density has to be calculated self-consistently, $j_c$
and $n_c$ are the angular-moment and the principal quantum number of the core electrons, and $G_c$ and
$F_c$ are the upper and the lower radial components of the wave function. The CH potential plays a
special role for alkaline ions since a well-defined part of the screening effects can be accounted
for by perturbing the first-order QED corrections with this potential (see, {\it e.g.},
Ref.~\cite{blundell:93:a}). Owing to this property, the CH potential was frequently employed for
an approximate treatment of the screening of QED effects
\cite{blundell:93:a,cheng:93,sapirstein:01:lamb}; a similar potential (without self-consistency)
was used in Refs.~\cite{indelicato:91:tca,indelicato:01}.

The second model potential is based on results of the density-functional theory (DFT) and referred
to as the Kohn-Sham (KS) potential. The local potential derived from DFT is given by
\cite{kohn:65,cowan,sapirstein:02:lamb}
\begin{eqnarray}\label{eq6}
    V_{\rm DFT}(r) &=& V_{\rm nuc}(r) + \alpha \int_0^{\infty} dr\pr \frac1{r_>}\,
        \rho_t^{s.c.}(r\pr)
 \nonumber \\ &&
     {} - x_{\alpha}\, \frac{\alpha}{r}\,
          \left[ \frac{81}{32\, \pi^2}\,r\,\rho_t^{s.c.}(r) \right]^{1/3}  \,,
\end{eqnarray}
where $\rho_t(r) = \rho_c(r)+\rho_v(r)$ is the total charge density, $\rho_v(r) =
G_v^2(r)+F_v^2(r)$ is the charge density of the valence electron, and $x_{\alpha} \in [0,1]$ is a
parameter. This potential has a non-physical limit at large $r$ and need, therefore, to be
corrected \cite{latter:55} to yield
\begin{equation}\label{eq7}
   V_{\rm DFT}(r)   \to -\frac{\alpha (Z-N_c)}{r}\,\ \ \ \mbox {\rm as}\ \  r\to\infty\,,
\end{equation}
where $N_c$ the number of the core electrons. In our calculations, we smoothly restore the correct
asymptotic behavior of the potential by adding a damping exponent. The result is
\begin{eqnarray}\label{eq8}
    V_{\rm KS}(r) &=& V_{\rm nuc}(r) + \alpha \int_0^{\infty} dr\pr \frac1{r_>}\,
        \rho_t^{s.c.}(r\pr)
 \nonumber \\ &&
        - x_{\alpha}\, \frac{\alpha}{r}\,
          \left[ \frac{81}{32 \pi^2}\,r\,\rho^{s.c.}_t(r) \right]^{1/3}
      - \frac{\alpha}{r}\,\left(1- e^{-A r^2} \right)
          \,,\nonumber \\
\end{eqnarray}
where the parameter $A$ has to be sufficiently small in order not to change the original DFT
potential significantly in the region of interest, but sufficiently large in order to restore the
proper asymptotic at the cavity radius. (In actual calculations, we put our system in a cavity with
the typical radius $R = 80/Z$~a.u.) The following value for the parameter $A$ was employed,
\begin{equation}\label{eq9}
    A = \frac1{100}\, \frac{(\Za)^2}{(n_r+\gamma)^2+(\Za)^2}\,,
\end{equation}
where $\gamma = \sqrt{\kappa_v^2-(\Za)^2}$ and $\kappa_v$ and $n_r$ are the Dirac quantum number
and the radial quantum number of the valence state $v$, respectively. The parameter $x_{\alpha}$
was set to be $x_{\alpha} = 2/3$ \cite{kohn:65,sapirstein:02:lamb}, if not specified otherwise.

%%%%%%%%%%%%%%%%%%%%%%%%%%%%%%%%%%%%%%%%%%%%%%%%%%%%%%%%%%%%%%%%%%%%%%%%
%%%%%
%%%%%
%%%%%%%%%%%%%%%%%%%%%%%%%%%%%%%%%%%%%%%%%%%%%%%%%%%%%%%%%%%%%%%%%%%%%%%
\begin{figure*}
\centerline{
\resizebox{0.98\textwidth}{!}{%
  \includegraphics{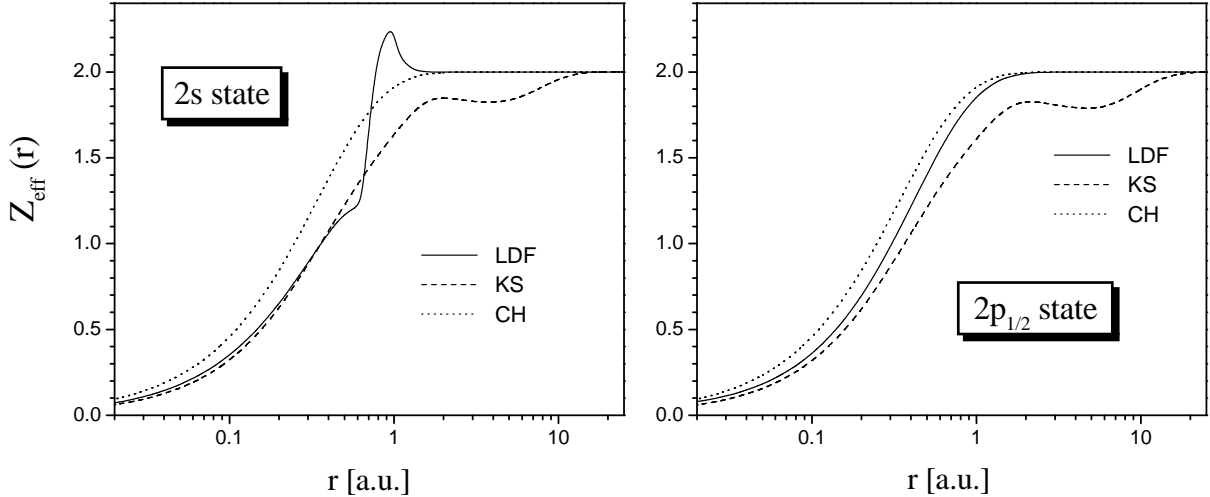}
}}
 \caption{ Radial dependence of the effective nuclear charge $Z_{\rm eff}(r)$ of different
screening potentials for the $(1s)^2\,2s$ and $(1s)^2\,2p_{1/2}$ states of lithium.
 \label{fig:potentials}}
\end{figure*}

The third model potential is obtained from the numerical solutions of the standard Dirac-Fock
(DF) problem; it is referred to as the local Dirac-Fock (LDF) potential. The scheme for
construction of this potential was developed previously in
Refs.~\cite{shabaev:05:prl,shabaev:05:pra}. For completeness, we reproduce it here, indicating some
modifications introduced in this work.

The solution of the radial DF equations for the valence state $v$ (achieved in this work with help
of the GRASP package \cite{parpia:96}) provides us with the eigenvalue $\vare_v^{\rm DF}$ and the
upper and the lower radial components denoted by  $G_v^{\rm DF}$ and $F_v^{\rm DF}$, respectively.
Let us now imagine that they are the eigenvalue and the solutions of the radial Dirac equations
with a local potential $V_v$,
\begin{align} \label{eq10}
  -\left(\frac{d}{dr}-\frac{\kappa_v}{r}\right)\, F_v^{\rm DF}(r) +V_v&(r)  \,G_v^{\rm DF}(r)
   \nonumber \\
    &  = (\vare_v^{\rm DF}-m)\,G_v^{\rm DF}(r) \,,
    \\ \label{eq11}
    \left(\frac{d}{dr}+\frac{\kappa_v}{r}\right)\, G_v^{\rm DF}(r) +V_v&(r)  \,F_v^{\rm DF}(r)
   \nonumber \\
   &  = (\vare_v^{\rm DF}+m)\,F_v^{\rm DF}(r)\,.
\end{align}
Generally speaking, one cannot find a local potential $V_v$ exactly satisfying both radial
equations. We can try, however, to solve this problem approximately. For the states $v$ with the wave
function without nodes [in our case, such are the $2p_{1/2}$ and $2p_{3/2}$ states], we obtain our
local potential by multiplying Eqs.~(\ref{eq10}) and (\ref{eq11}) by $G_v^{\rm DF}$ and $F_v^{\rm
DF}$, respectively, adding them together, and then inverting the equation with respect to $V_v$.
The resulting potential is
\begin{eqnarray}\label{eq12}
    V_{{\rm LDF},v}^{(1)}(r) &=& \vare_v^{\rm DF} + \frac{G_v^{\rm DF}}{\rho_v^{\rm DF}}
       \left(\frac{d}{dr}-\frac{\kappa_v}{r}\right) F_v^{\rm DF}
 \nonumber \\ &&
         - \frac{F_v^{\rm DF}}{\rho_v^{\rm DF}}
       \left(\frac{d}{dr}+\frac{\kappa_v}{r}\right) G_v^{\rm DF}
 \nonumber \\ &&
       - m\,\frac{\left(G_v^{\rm DF}\right)^2-\left(F_v^{\rm DF}\right)^2}{\rho_v^{\rm DF}}\,,
\end{eqnarray}
where $v$ is the valence state with $n_r=0$. In the case when the valence wave function has nodes
($n_r>0$), the potential (\ref{eq12}) has to be modified since the density $\rho_v^{\rm DF}$
(although positively defined everywhere) vanishes in the nonrelativistic limit at the nodes of the
upper component, making the potential to be nearly singular at these points.

In order to construct a smooth version of the potential (\ref{eq12}) for $n_r >0$, we define the
density $\overline{\rho}_v$ that has an admixture of the density of core states with the same value
of $\kappa$ as the valence state,
\begin{equation}\label{eq13}
  \overline{\rho}_v = \sum_{n \leq n_v,\  \kappa = \kappa_v} w_{\kappa,n}\,\rho^{\rm DF}_{\kappa,n}\,,
  \ \ \ \sum_{n \leq n_v,\  \kappa = \kappa_v} w_{\kappa,n} = 1\,.
\end{equation}
By a proper choice of the weights $w_{\kappa,n}$, the density $\overline{\rho}_v$ can be made to
behave more smoothly than $\rho_v$. At the same time, the admixture of the core density should be
small in order not to disturb the potential outside of the vicinity of the nodes. We achieve this
by choosing the weights to be dependent on $r$,
\begin{equation}\label{eq14}
    w_{\kappa,n} \propto \frac{\rho^{\rm DF}_{\kappa_v,n_v}(r\to 0)}{\rho^{\rm DF}_{\kappa,n}(r\to 0)}\,
       \sum_{i=1}^{n_r} \exp\left[-10\,(\Za)^2\,(r-r_i)^2) \right]
\end{equation}
for $n < n_v$, and $w_{\kappa_v,n_v} \propto 1$, where $r_i$ are the nodes of the upper component.
The choice of weights $w_{\kappa,n}$ in the form (\ref{eq14}) is the only difference of the present
LDF potential from the one constructed in Refs.~\cite{shabaev:05:prl,shabaev:05:pra}.

In order to build a potential with the density $\overline{\rho}_v$, we first define a new
function $X_v$, representing the potential (\ref{eq12}) in the form
\begin{eqnarray}\label{eq15}
    V_{{\rm LDF},v}^{(1)}(r) &=& V_{\rm nuc}(r)
      + \alpha \int_0^{\infty} dr\pr \frac1{r_>}\, \rho_c^{\rm DF}(r\pr)
 \nonumber \\ &&
        +  \frac1{r\, \rho_v^{\rm DF}(r)}\, X_v(r)\,.
\end{eqnarray}
$X_v$ is a smooth function also for $n_r>0$. Analogously, we introduce the functions $X_{\kappa,n}$
for each core state in Eq.~(\ref{eq13}). Finally, we define our local potential as
\begin{eqnarray}\label{eq16}
    V_{{\rm LDF},v}(r) &=& V_{\rm nuc}(r)
      + \alpha \int_0^{\infty} dr\pr \frac1{r_>}\, \rho_c^{\rm DF}(r\pr)
 \nonumber \\ &&
        +  \frac1{r\, \overline{\rho}_v(r)}\,\sum_{n \leq n_v,\  \kappa = \kappa_v} w_{\kappa,n}\, X_{\kappa,n}(r)\,.
\end{eqnarray}
Obviously, $V_{{\rm LDF},v} = V_{{\rm LDF},v}^{(1)}$ for $n_r = 0$.

It is of interest to compare the radial dependence of different screening potentials defined above.
In Fig.~\ref{fig:potentials}, we present such comparison for the $(1s)^2\,2s$ and
$(1s)^2\,2p_{1/2}$ states of lithium, expressing the potentials in terms the effective nuclear
charge
\begin{equation}\label{efnucch}
 Z_{\rm eff}(r) =  \frac{r}{\alpha}\,[V_{\rm eff}(r)-V_{\rm nuc}(r)] \,,
\end{equation}
where $V_{\rm eff}(r)$ is one of the LDF, KS, and CH potentials. Further discussion of different
potentials will be postponed until the next section, where we will be able to compare the
convergence of the perturbative expansions based on them.

%%%%%%%%%%%%%%%%%%%%%%%%%%%%%%%%%%%%%%%%%%%%%%%%%%%%%%%%%%%%%%%%%%%%%%%%%%%%%%%%%%%%%%%%%%%%%%
%
%%%%%%%%%%%%%%%%%%%%%%%%%%%%%%%%%%%%%%%%%%%%%%%%%%%%%%%%%%%%%%%%%%%%%%%%%%%%%%%%%%%%%%%%%%%%%%

\section{MBPT treatment}
\label{sec:mbpt}

The MBPT picture can be obtained from the full QED theory by applying the Breit approximation to
the operator of the electron-electron interaction $I(\omega)\to I = I^{\rm Coul}+I^{\rm Breit}$, as
indicated by Eqs.~(\ref{eq3})-(\ref{eq3a}). Using the Breit approximation in the second and higher
orders of perturbation theory, it should be taken into account that a summation over the
negative-energy part of the Dirac spectrum may lead to large spurious effects \cite{sucher:84}.
Consistent treatment of the negative-energy states in such situations can be performed only within QED
and the common approach is to restrict the summations to the positive-energy part of the Dirac
spectrum only.

General formulas for the MBPT corrections to the energy for systems with a single electron outside
a closed core are known and can be found in Ref.~\cite{blundell:87:adndt}. It is our intention,
however, to obtain a different representation for them in order to make explicit connections with the
corresponding formulas in full QED.

As the zeroth-order approximation, we take the eigenvalues and the solutions of the Dirac equation
with the effective potential $V_{\rm eff}(r) = V_{\rm nuc}(r)+ U(r)$, where $U(r)$ is the model
potential that accounts for a part of the electron-electron interaction. Since we intend to
incorporate the MBPT results into {\em ab initio} QED calculations, the model potential $U$ should be
a {\em local} one, thus excluding the possibility to use the Dirac-Fock potential, which is the
standard choice in traditional many-body calculations.

In the following, we assume the electron configuration to be of the form $(1s)^2\,v$, where $v$ is
the valence electron and $1s$ is a core electron (also denoted by $c$ in formulas below). The
contributions due to the core-core interaction will be excluded from consideration as they do not
influence transition energies; corrections obtained in this way can be identified as contributions
to the ionization energy of the valence electron.

The zeroth-order energy $E^{(0)}$ is defined as
\begin{equation}\label{eqII1}
    E^{(0)} = \vare_v-m\,,
\end{equation}
where $\vare_v$ is the Dirac eigenvalue corresponding to the valence state $v$. Corrections to the
lowest-order energy are treated by perturbation theory. Since the model potential $U$ is included
into the zeroth-order approximation, we have to account for the ``residual" interaction $(-U)$ in
each order of perturbation theory and the perturbative expansion goes both in the electron-electron
interaction and in $(-U)$ (see Ref.~\cite{sapirstein:98:rmp} for details).

We will denote corrections to the energy by $\Delta E^{(i,j|k)}$, where the index $i$ indocates the
order of the correction with respect to the electron-electron interaction, $j$ denotes the order in
$U$, and $k = i+j$ is the total order of perturbation theory. The following shorthand notations
will be used throughout the paper: $I_{{a}{b}{c}{d}} = \lbr ab|I|cd\rbr\,$, $U_{{a}{b}} = \lbr
a|U|b\rbr\,$. By the symbols $\delta^{(i)}$ we will denote the $i$th-order (in $U$) perturbation of
the wave function. The first two corrections are
\begin{equation}\label{eqII2}
    |\delta^{(1)}a\rbr = \sum_{n}{}^{^{\prime}}
    \frac{U_{{n}{a}}\,|n\rbr}{\vare_a-\vare_n}\,,
\end{equation}
\begin{eqnarray}\label{eqII3}
    |\delta^{(2)}a\rbr &=& \sum_{n}{}^{^{\prime}}
    \frac{\U{n}{\delta^{(1)}a}\,|n\rbr}{\vare_a-\vare_n}
 - \U{a}{a}\,\sum_{n}{}^{^{\prime}}
    \frac{U_{{n}{a}}\,|n\rbr}{(\vare_a-\vare_n)^2}
    \nonumber \\ &&
 -\frac12\,|a\rbr\, \lbr \delta^{(1)}a|\delta^{(1)}a\rbr\,,
\end{eqnarray}
where the prime on a sum denotes that the terms with the vanishing denominator should be omitted.
We found it important to keep the summation over the complete Dirac spectrum in the above
corrections to the wave functions, not excluding the negative-energy states as is customary in MBPT
calculations. Similar conclusion was previously drawn in Ref.~\cite{sapirstein:99}, where it was
demonstrated that inclusion of some negative-energy states drastically reduces the potential
dependence of MBPT results for He-like ions.

For the MBPT corrections that do not depend on the model potential, we obtain
\begin{equation}\label{eqII4}
   \Delta E^{(1,0|1)}_{\rm MBPT} = \sum_{\mu_c}\sum_P (-1)^P\,\I{Pv}{Pc}{v}{c} \,,
\end{equation}
\begin{widetext}
\begin{eqnarray}\label{eqII5}
    \Delta E^{(2,0|2)}_{\rm MBPT} &=& \sum_{\mu_c}\sum_P
    (-1)^P\,\sum_{n_1n_2}{}^{^{\prime}}{}^{^{(+)}}
        \frac{\I{Pv}{Pc}{n_1}{n_2}\,\I{n_1}{n_2}{v}{c}}{\vare_c+\vare_v-\vare_{n_1}-\vare_{n_2}}
%    \nonumber \\ &&
    + \sum_{PQ}(-1)^{P+Q}\, \sum_{n}{}^{^{\prime}}{}^{^{(+)}}
      \frac{\I{P2}{P3}{n}{Q3}\,\I{P1}{n}{Q1}{Q2}}{\vare_{Q1}+\vare_{Q2}-\vare_{P1}-\vare_{n}}\,,
%      \nonumber \\
\end{eqnarray}
\begin{eqnarray}\label{eqII6}
 \Delta E^{(3,0|3)}_{\rm MBPT} &=&
    \sum_{\mu_c}\, \sum_P (-1)^P\,
     \sum_{n_1\ldots n_4}{\!\!}^{^{(+)}}\,
         \Xi_1 \,\frac{\I{Pv}{Pc}{n_1}{n_2}\, \I{n_1}{n_2}{n_3}{n_4}\,
         \I{n_3}{n_4}{v}{c}}
         {(\vare_c+\vare_v-\vare_{n_1}-\vare_{n_2})(\vare_c+\vare_v-\vare_{n_3}-\vare_{n_4})}
   \nonumber \\ &&
{} +  \sum_{PQ} (-1)^{P+Q}\,   \sum_{n_1 n_2 n_3}{\!\!\!}^{^{(+)}}\, \Xi_1 \,\left[
         \frac{2\,\I{P2}{P3}{n_1}{Q3}\,
    \I{P1}{n_1}{n_2}{n_3}\, \I{n_2}{n_3}{Q1}{Q2}}
          {(\vare_{Q1}+\vare_{Q2}-\vare_{P1}-\vare_{n_1})(\vare_{Q1}+\vare_{Q2}-\vare_{n_2}-\vare_{n_3})}
 \right. \nonumber \\ &&
{}    + \frac{\I{P1}{P2}{n_1}{n_2}\,
    \I{n_2}{P3}{n_3}{Q3}\, \I{n_1}{n_3}{Q1}{Q2}}
          {(\vare_{P1}+\vare_{P2}-\vare_{n_1}-\vare_{n_2})(\vare_{Q1}+\vare_{Q2}-\vare_{n_1}-\vare_{n_3})}
%       \nonumber \\ &&
       \left.
{}+         \frac{\I{P2}{P3}{n_1}{n_2}\,
    \I{P1}{n_1}{n_3}{Q2}\, \I{n_3}{n_2}{Q1}{Q3}}
          {(\vare_{P2}+\vare_{P3}-\vare_{n_1}-\vare_{n_2})(\vare_{Q1}+\vare_{Q3}-\vare_{n_2}-\vare_{n_3})}
\right]\,,
 \nonumber \\
\end{eqnarray}
\end{widetext}
 where $\mu_c$ denotes the momentum projection of the core electron, $P$ and
$Q$ are the permutation operators [$(PvPc) = (vc)$, $(cv)$], and the sign ``$(+)$" on a sum
indicates that the summation is performed over the positive-energy part of the Dirac spectrum only.
For the three-electron corrections, the numbers 1, 2, and 3 numerate the electrons in the
configuration (in arbitrary order). The operator $\Xi_1$ acts on energy denominators $\Delta_1$,
$\Delta_2$ and is defined as
\begin{eqnarray} \label{eqII7}
\Xi_1 \, \frac{X}{\Delta_1\,\Delta_2} = \left\{
   \begin{array}{cl}
 \displaystyle       \frac{ X}{ \Delta_1\,\Delta_2}\,,
        & \mbox{if}\         \Delta_1\ne0\,,\Delta_2\ne0\,,\\[0.2cm]
        \displaystyle -\frac{ X}{2\,\Delta_1^2}\,, &\mbox{if}\
                             \Delta_1\ne0\,,\Delta_2=0\,,\\[0.2cm]
        \displaystyle  -\frac{X}{2\,\Delta_2^2}\,, &\mbox{if}\
                             \Delta_1=0\,,\Delta_2\ne0\,,\\[0.2cm]
        0\,, &\mbox{if}\
                             \Delta_1=0\,,\Delta_2=0\,.\\
   \end{array}
   \right.
\end{eqnarray}

The corrections of the form $\Delta E^{(0,i|i)}$ are just the $i$th-order [in $(-U)$] perturbations
of the Dirac energy,
\begin{equation} \label{eqII8}
    \Delta E^{(0,1|1)} = -\U{v}{v}\,,
\end{equation}
\begin{equation} \label{eqII9}
    \Delta E^{(0,2|2)} = \U{\delta^{(1)}v}{v}\,,
\end{equation}
\begin{equation} \label{eqII10}
  \Delta E^{(0,3|3)} = -\U{\delta^{(1)}v}{\delta^{(1)}v}+
                \U{v}{v}\,\lbr\delta^{(1)}v|\delta^{(1)}v\rbr\,.
\end{equation}

The corrections of the form $\Delta E^{(1,i|i+1)}_{\rm MBPT}$ are the $i$th-order
perturbations of the one-photon exchange contribution $\Delta E^{(1,0|1)}_{\rm
MBPT}$. They are given by
\begin{equation} \label{eqII11}
    \Delta E^{(1,1|2)}_{\rm MBPT} = -2 \sum_{\mu_c}\sum_P (-1)^P \Bigl[
    \I{Pv}{Pc}{\delta^{(1)}v}{c}+\I{Pv}{Pc}{v}{\delta^{(1)}c} \Bigr] \,,
\end{equation}
\begin{widetext}
\begin{eqnarray} \label{eqII12}
 \Delta E^{(1,2|3)}_{\rm MBPT}  &=&
    \sum_{\mu_c}\, \sum_P (-1)^P\, \Bigl[
  2\,\I{Pv}{Pc}{\delta^{(2)}v}{c} +
  2\,\I{Pv}{Pc}{v}{\delta^{(2)}c} + 2\,\I{Pv}{Pc}{\delta^{(1)}v}{\delta^{(1)}c}
  \nonumber \\ &&
  + \I{\delta^{(1)}Pv}{Pc}{\delta^{(1)}v}{c}
  + \I{\delta^{(1)}Pv}{Pc}{v}{\delta^{(1)}c}
  + \I{Pv}{\delta^{(1)}Pc}{\delta^{(1)}v}{c}
  + \I{Pv}{\delta^{(1)}Pc}{v}{\delta^{(1)}c}
  \Bigr]\,.
\end{eqnarray}

Finally, $\Delta E^{(2,1|3)}_{\rm MBPT}$ is the first-order correction to the
two-photon exchange contribution $\Delta E^{(2,0|2)}_{\rm MBPT}$. It consists of 3
parts that arise as perturbations of the wave functions (``wf"), binding energies
(``en"), and propagators (``pr"),
\begin{equation} \label{eqII13}
\Delta E^{(2,1|3)}_{\rm MBPT} = \Delta E^{(2,1|3)}_{\rm wf}+ \Delta E^{(2,1|3)}_{\rm en}+
\Delta E^{(2,1|3)}_{\rm pr}\,.
\end{equation}
The corresponding corrections are given by
\begin{eqnarray} \label{eqII14}
    \Delta E^{(2,1|3)}_{\rm wf} &=& -2\,\sum_{\mu_c}\sum_P
    (-1)^P\,
    \sum_{n_1n_2}{}^{^{\prime}}{}^{^{(+)}}
        \frac{\I{Pv}{Pc}{n_1}{n_2}\, \left[ \I{n_1}{n_2}{\delta^{(1)}v}{c}+ \I{n_1}{n_2}{v}{\delta^{(1)}c}\right]}
                      {\vare_c+\vare_v-\vare_{n_1}-\vare_{n_2}}
    \nonumber \\ &&
{} - 2\,  \sum_{PQ}(-1)^{P+Q}\,
       \sum_{n}{}^{^{\prime}}{}^{^{(+)}}
       \frac{\I{P2}{P3}{n}{Q3}\, \bigl[ \I{\delta^{(1)}P1}{n}{Q1}{Q2}+
      \I{P1}{n}{\delta^{(1)}Q1}{Q2}+ \I{P1}{n}{Q1}{\delta^{(1)}Q2}\bigr]}
      {\vare_{Q1}+\vare_{Q2}-\vare_{P1}-\vare_{n}}\,,
      \nonumber \\
\end{eqnarray}
\begin{eqnarray} \label{eqII15}
    \Delta E^{(2,1|3)}_{\rm en} &=&
 (\U{v}{v}+\U{c}{c})\, \sum_{\mu_c}\sum_P
    (-1)^P\,\sum_{n_1n_2}{}^{^{\prime}}{}^{^{(+)}}
        \frac{\I{Pv}{Pc}{n_1}{n_2}\,\I{n_1}{n_2}{v}{c}}{(\vare_c+\vare_v-\vare_{n_1}-\vare_{n_2})^2}
    \nonumber \\ &&
  {}  + \sum_{PQ}(-1)^{P+Q}\, (\U{Q1}{Q1}+\U{Q2}{Q2}-\U{P1}{P1})\,
    \sum_{n}{}^{^{\prime}}{}^{^{(+)}}
      \frac{\I{P2}{P3}{n}{Q3}\,\I{P1}{n}{Q1}{Q2}}{(\vare_{Q1}+\vare_{Q2}-\vare_{P1}-\vare_{n})^2}\,,
      \nonumber \\
\end{eqnarray}
\begin{eqnarray} \label{eqII16}
    \Delta E^{(2,1|3)}_{\rm pr} &=&
 -\sum_{\mu_c}\sum_P
    (-1)^P\,\sum_{n_1n_2n_3}{\!\!}^{^{(+)}}\, \Xi_2\,
        \left[ \frac{\I{Pv}{Pc}{n_1}{n_2}\,\U{n_1}{n_3}\,\I{n_3}{n_2}{v}{c}}{(\vare_c+\vare_v-\vare_{n_1}-\vare_{n_2})
              (\vare_c+\vare_v-\vare_{n_3}-\vare_{n_2})}
    \right. \nonumber \\ && \left.
\ \ \ \ \ \ \ \ \ \ \ \ \ \ \ \ \ \ \ \ \ \ \ \ \ \ \ \ \ \ \ \ \ \
\ \ \ \ \ \ \ \ \ \ \ \ \
 +\frac{\I{Pv}{Pc}{n_1}{n_2}\,\U{n_2}{n_3}\,\I{n_1}{n_3}{v}{c}}{(\vare_c+\vare_v-\vare_{n_1}-\vare_{n_2})
              (\vare_c+\vare_v-\vare_{n_1}-\vare_{n_3})}
    \right]
    \nonumber \\ &&
    - \sum_{PQ}(-1)^{P+Q}\, \sum_{n_1n_2}{}^{^{(+)}}\,
    \Xi_2\,  \frac{\I{P2}{P3}{n_1}{Q3}\,\U{n_1}{n_2}\,\I{P1}{n_2}{Q1}{Q2}}
    {(\vare_{Q1}+\vare_{Q2}-\vare_{P1}-\vare_{n_1})(\vare_{Q1}+\vare_{Q2}-\vare_{P1}-\vare_{n_2})}\,,
\end{eqnarray}
 where the operator $\Xi_2$ acts on energy denominators $\Delta_1$,
$\Delta_2$ as following:
\begin{eqnarray} \label{eqII17}
\Xi_2 \, \frac{X}{\Delta_1\,\Delta_2} = \left\{
   \begin{array}{cl}
 \displaystyle       \frac{ X}{ \Delta_1\,\Delta_2}\,,
        & \mbox{if}\         \Delta_1\ne0\,,\Delta_2\ne0\,,\\[0.2cm]
        \displaystyle -\frac{ X}{\Delta_1^2}\,, &\mbox{if}\
                             \Delta_1\ne0\,,\Delta_2=0\,,\\[0.2cm]
        \displaystyle  -\frac{X}{\Delta_2^2}\,, &\mbox{if}\
                             \Delta_1=0\,,\Delta_2\ne0\,,\\[0.2cm]
        0\,, &\mbox{if}\
                             \Delta_1=0\,,\Delta_2=0\,.\\
   \end{array}
   \right.
\end{eqnarray}

Collecting together all corrections to a given order of perturbation theory, we obtain
\begin{eqnarray} \label{eqII18}
 \Delta E^{(1)}_{\rm MBPT} &=& \Delta E^{(1,0|1)}_{\rm MBPT}+\Delta E^{(0,1|1)} \,, \\
                 \label{eqII18a}
 \Delta E^{(2)}_{\rm MBPT} &=& \Delta E^{(2,0|2)}_{\rm MBPT}+\Delta E^{(1,1|2)}_{\rm MBPT}+\Delta E^{(0,2|2)} \,, \\
 \Delta E^{(3)}_{\rm MBPT} &=& \Delta E^{(3,0|3)}_{\rm MBPT}+\Delta E^{(2,1|3)}_{\rm MBPT}
% \nonumber \\ &&
     +\Delta E^{(1,2|3)}_{\rm MBPT}+\Delta E^{(0,3|3)} \,.
     \label{eqII19}
\end{eqnarray}

The expressions for the second- and third-order corrections presented above contain products of the
operators $I$ and, therefore, include the Breit interaction to the second and even to the third
order. Keeping the Breit interaction to the second order produces contributions of the same order
as QED effects omitted [for the two-photon exchange, the order is $\alpha^2(\Za)^3$] and thus
cannot be regarded as ultimately wrong. However, this contribution can be shown to originate from
the region of virtual excitation energies of order of the electron mass, where the Breit
approximation is no longer valid. We thus consider it to be more correct to treat the Breit
interaction as a first-order perturbation only. In our actual calculations, we expand each $I$ into
a sum of the Coulomb and Breit parts and keep terms with the Breit interaction up to the first
order only.

The MBPT results (\ref{eqII18})-(\ref{eqII19}) can easily be improved by accounting for the
one-photon exchange correction rigorously, using the exact expression for the operator of the
electron-electron interaction (\ref{eq1}). In this case, approximate formulas (\ref{eqII4}),
(\ref{eqII11}), and (\ref{eqII12}) should be replaced by the exact ones
\begin{equation}\label{eqII20}
    \Delta E^{(1,0|1)} =  \sum_{\mu_c}\sum_P (-1)^P
    \I{Pv}{Pc}{v}{c}(\Delta_{Pc\,c}) \,,
\end{equation}
\begin{eqnarray}\label{eqII21}
    \Delta E^{(1,1|2)} &=& -2 \sum_{\mu_c}\sum_P (-1)^P \Bigl[
    \I{Pv}{Pc}{\delta^{(1)}v}{c}(\Delta_{Pc\,c})
       +\I{Pv}{Pc}{v}{\delta^{(1)}c}(\Delta_{Pc\,c}) \Bigr]
%  \nonumber \\ && {}
   + \Delta^{(1)}_{vc}\, \sum_{\mu_c} I^{\prime}_{cvvc}(\Delta_{vc})
       \,,
\end{eqnarray}
\begin{eqnarray} \label{eqII22}
 \Delta E^{(1,2|3)}  &=&
    \sum_{\mu_c}\, \sum_P (-1)^P\, \Bigl[
  2\,\I{Pv}{Pc}{\delta^{(2)}v}{c}(\Delta_{Pc\,c}) +
  2\,\I{Pv}{Pc}{v}{\delta^{(2)}c}(\Delta_{Pc\,c})
  + 2\,\I{Pv}{Pc}{\delta^{(1)}v}{\delta^{(1)}c}(\Delta_{Pc\,c})
  \nonumber \\ &&
  + \I{\delta^{(1)}Pv}{Pc}{\delta^{(1)}v}{c}(\Delta_{Pc\,c})
  + \I{\delta^{(1)}Pv}{Pc}{v}{\delta^{(1)}c}(\Delta_{Pc\,c})
  + \I{Pv}{\delta^{(1)}Pc}{\delta^{(1)}v}{c}(\Delta_{Pc\,c})
  + \I{Pv}{\delta^{(1)}Pc}{v}{\delta^{(1)}c}(\Delta_{Pc\,c})
  \Bigr]
  \nonumber \\ &&
-2\, \Delta^{(1)}_{vc}\, \sum_{\mu_c} \Bigl[
    \DI{c}{v}{\delta^{(1)}v}{c}(\Delta_{vc})
       +\DI{c}{v}{v}{\delta^{(1)}c}(\Delta_{vc}) \Bigr]
-\frac12\,\Delta^{(1)^2}_{vc}\, \sum_{\mu_c} I^{\prime\prime}_{cvvc}(\Delta_{vc})
-\Delta^{(2)}_{vc}\, \sum_{\mu_c} I^{\prime}_{cvvc}(\Delta_{vc})
  \,,
\end{eqnarray}

\end{widetext}
where $I_{abcd}(\Delta) = \lbr ab|I(\Delta)|cd\rbr$ and $I(\Delta)$ is the exact operator of the
electron-electron interaction given by Eq.~(\ref{eq1}). Other notations are: $\Delta_{ab} =
\vare_a-\vare_b$, $\Delta^{(1)}_{vc} = \U{v}{v}-\U{c}{c}$,  $\Delta^{(2)}_{vc} =
\U{\delta^{(1)}v}{v}-\U{\delta^{(1)}c}{c}\,$, and $I^{\prime}(\Delta) = \left. d/(d
\vare)\,I(\vare) \right|_{\vare=\Delta}$.

The resulting set of corrections to $E^{(0)}$ is given by
\begin{eqnarray} \label{eqII23}
\Delta E^{(1)} &=& \Delta E^{(1,0|1)}+\Delta E^{(0,1|1)} \,, \\
   \label{eqII23a}
\widetilde{  \Delta E}^{(2)}_{\rm MBPT} &=& \Delta E^{(2,0|2)}_{\rm MBPT}+\Delta E^{(1,1|2)}+\Delta E^{(0,2|2)} \,, \\
\widetilde{  \Delta E}^{(3)}_{\rm MBPT} &=& \Delta E^{(3,0|3)}_{\rm MBPT}+\Delta
E^{(2,1|3)}_{\rm MBPT}
 \nonumber \\ &&
     +\Delta E^{(1,2|3)}+\Delta E^{(0,3|3)} \,.
     \label{eqII24}
\end{eqnarray}
Now we have the exact expression for the first-order correction $\Delta E^{(1)}$, whereas the
second-order correction still contains the two-photon exchange part calculated within the MBPT
approximation only. The rigorous QED treatment of the correction $\Delta E^{(2,0|2)}$ will be
presented in the next section. The three-photon exchange contribution is more difficult to evaluate
rigorously; presently it can be addressed to only within MBPT or other methods based on the Breit
approximation.

We now discuss some details of our numerical evaluation of the MBPT corrections. The complete set
of Dirac eigenstates for a given local potential $V_{\rm eff}(r)$ is generated by means of the
dual-kinetic-balance basis-set method \cite{shabaev:04:DKB} with basis functions constructed from
$B$ splines. The computationally most intensive part of the calculation is the third-order
correction $\Delta E_{\rm MBPT}^{(3,0|3)}$, particularly the first term in Eq.~(\ref{eqII6}) that
involves a quadruple sum over intermediate states. The summation over magnetic substates was
performed numerically for each coupling by summing combinations of the Clebsch-Gordan coefficients.
It is possible to derive closed expressions for such sums (as was done for the third-order Coulomb
exchange in Ref.~\cite{johnson:88:b}). However, since it does not significantly influence the
performance of our code, we prefer to do this summation numerically.

The main problem with the evaluation of the first term in Eq.~(\ref{eqII6}) is that it involves two
independent (and unrestricted) partial-wave summations over the Dirac angular momentum of
intermediate states, which we denote as $\kappa_1$ and $\kappa_2$. Two other summations (over
$\kappa_3$ and $\kappa_4$) become finite after taking into account selection rules of the Racah
algebra. Our approach to the evaluation of the double partial-wave expansion over $\kappa_1$ and
$\kappa_2$ is similar to the one used in the previous calculation of the two-loop self-energy
correction \cite{yerokhin:03:prl,yerokhin:03:epjd}. First, we construct a matrix of elements with
fixed values of $|\kappa_1|$ and $|\kappa_2|$: $X_{|\kappa_1|,|\kappa_2|}$, $|\kappa_1| = 1,\ldots$
and $|\kappa_2| = 1,\ldots$. Then, we extrapolate to infinity the partial sums along the diagonals
of the matrix $X_{|\kappa_1|,|\kappa_2|}$, taking typically of about 5 terms for each diagonal.
Finally, we extrapolate to infinity the sequence of the sums of $i$th subdiagonals of the matrix,
taking typically of about 7 terms. Significant numerical cancellations should be mentioned that
arise in our numerical evaluation of the third-order correction in the low-$Z$ region. This leads to
necessity to employ a sufficiently large basis set in order to achieve a proper convergence. In
actual calculations, we used a basis set with the number of positive-energy states varying from 50
in the low-$Z$ region to 40 in the high-$Z$ region.

An important question that arises in actual calculations is how one should interpret the condition
of the vanishing of the denominator, which is used in order to exclude terms from the summation in
Eqs.~(\ref{eqII5}) {\em etc.} and to define the case in Eqs.~(\ref{eqII7}) and (\ref{eqII17}). The
point is that, for $v=2p_J$, an intermediate state with the same orbital momentum $l$, namely
$2p_{J\pr}$, appears in the Dirac spectrum, whose energy is separated from $\vare_v$ by
relativistic effects only, $\delta = \vare_{2p_{3/2}}-\vare_{2p_{1/2}}\propto (\Za)^2$. As a
result, very small energy denominators of order $\delta^2$ may appear in the expressions for the
third-order energy, which would lead to huge numerical cancellations in the low-$Z$ region.

The appearance of small denominators can be avoided if we take into account that there could be no
contributions to the energy of the state $(1s)^22p_{J}$ arising from the intermediate 3-electron
state $(1s)^22p_{J\pr}$ with a different total momentum $J\pr$. This means that the total
contribution of such intermediate states is zero, although the corresponding
terms can arise at intermediate stages of the calculation. We have verified this statement
numerically, which served as an additional cross-check of our numerical routine.
%It could be
%mentioned that a small non-zero contribution from such intermediate states was found for the
%correction $\Delta E_{\rm MBPT}^{(2,1|3)}$, which is noticeable for high $Z$ only and is probably
%due to incomplete treatment of the negative-energy states.
In the actual calculations, we define the energy denominator as vanishing if its absolute value is
smaller than the fine-structure difference $\delta$. This redefinition greatly facilitates the
numerical evaluation of the third-order correction in the low-$Z$ region.

%%%%%%%%%%%%%%%%%%%%%%%%%%%%%%%%%%%%%%%%%%%%%%%%%%%%%%%%%%%%%%%%%%%%%%%%%%%
%
%       Z=3
%
%%%%%%%%%%%%%%%%%%
\begin{table}[t]
\caption{MBPT contributions to the ionization potential of $n=2$ states of lithium
for different model potentials, in a.u.
 \label{tab:Li} }
\begin{ruledtabular}
\begin{tabular}{l....}
                &   \multicolumn{1}{c}{  Coul  }       &   \multicolumn{1}{c}{  CH }
                    &   \multicolumn{1}{c}{ KS}            &  \multicolumn{1}{c}{LDF} \\
\hline\\[-9pt]
$2s_{1/2}$ state: \\
   Dirac:      &     -1.1x2517   &     -0.1x8310   &     -0.2x4861   &     -0.1x9601   \\
   1-phot:     &      1.1x9370   &     -0.0x2159   &      0.0x5362   &     -0.0x0031   \\
   2-ph(MBPT): &     -0.2x5067   &      0.0x1158   &     -0.0x0334   &     -0.0x0184   \\
   3-ph(MBPT): &     -0.0x0838   &     -0.0x0939   &      0.0x0017   &      0.0x0001   \\
    Sum:       &     -0.1x9051   &     -0.2x0249   &     -0.1x9816   &     -0.1x9815   \\
   Exact:      &    \multicolumn{4}{c}{ $-0.198\,159\,72$} \\
 $2p_{1/2}$ state: \\
   Dirac:      &     -1.1x2517   &     -0.1x2704   &     -0.1x8537   &     -0.1x2867   \\
   1-phot:     &      1.4x0602   &     -0.0x0401   &      0.0x6161   &     -0.0x0109   \\
   2-ph(MBPT): &     -0.3x7129   &      0.0x0172   &     -0.0x0734   &     -0.0x0045   \\
   3-ph(MBPT): &     -0.0x2608   &     -0.0x0186   &      0.0x0120   &     -0.0x0009   \\
    Sum:       &     -0.1x1651   &     -0.1x3119   &     -0.1x2990   &     -0.1x3029   \\
   Exact:      &    \multicolumn{4}{c}{ $-0.130\,242\,69 $}\\
 $2p_{3/2}$ state: \\
   Dirac:      &     -1.1x2503   &     -0.1x2704   &     -0.1x8536   &     -0.1x2867   \\
   1-phot:     &      1.4x0571   &     -0.0x0401   &      0.0x6160   &     -0.0x0108   \\
   2-ph(MBPT): &     -0.3x7105   &      0.0x0172   &     -0.0x0734   &     -0.0x0046   \\
   3-ph(MBPT): &     -0.0x2617   &     -0.0x0187   &      0.0x0119   &     -0.0x0009   \\
    Sum:       &     -0.1x1654   &     -0.1x3120   &     -0.1x2991   &     -0.1x3029   \\
   Exact:      &     \multicolumn{4}{c}{ $-0.130\,241\,17 $}\\
\end{tabular}
\end{ruledtabular}
\end{table}

Let us now address the following questions: (i) what is the best local potential to use for the
perturbative expansion and (ii) how to estimate the residual electron-correlation effects? In order
to answer these questions, we turn to the most difficult Li-like atom to describe within MBPT, to
neutral lithium. For higher-$Z$ systems, the relative weight of the electron correlation with
respect to the binding energy gradually decreases, so one should expect that the perturbation 
expansion
converges faster. If we use the pure Coulomb potential in the zeroth-order approximation, the
perturbative-expansion parameter is just $1/Z$. It is no longer so for other potentials, but the
general tendency remains. Obviously, MBPT is not the best approach to use for very low-$Z$ systems.
Extremely accurate results are nowadays available for lithium; they can be used as a benchmark to
estimate the residual correlation effects in our calculations. The best results for lithium are
obtained within the method that starts with a variational solution of the three-body Schr\"odinger
problem with a fully correlated basis set and then adds relativistic and QED corrections term by
term as obtained within the $\Za$ expansion \cite{yan:98:prl,puchalski:06}.

Our numerical results for the ionization energy of the $n=2$ states of lithium are presented in
Table~\ref{tab:Li} for four potentials: the pure Coulomb (Coul), the core-Hartree, the Kohn-Sham,
and the local Dirac-Fock potentials. The entry labeled ``Dirac'' contains the zeroth-order energies
$E^{(0)}$; the next 3 lines contain the corrections given by Eqs.~(\ref{eqII23})-(\ref{eqII24}),
respectively. The ``exact'' results represent the sum of the nonrelativistic energy and the Breit
correction in the infinitely-heavy-nucleus limit; they are taken from Ref.~\cite{yan:98:prl}.

It is not at all unexpected that for lithium, the perturbation theory starting with the Coulomb
potential yields a very slowly converging series and that the results do not reproduce well the
``exact'' relativistic energies. It is perhaps more surprising that the CH potential, which gives a
reasonable value for the zeroth-order energy, does not nevertheless yield a well-converging
perturbative expansion. We have to conclude that the residual interaction is not small enough in
this case. Although the CH perturbative expansion becomes converging for higher values of $Z$, this
choice of the model potential does not look favorable when compared with the other potentials,
the KS and the LDF one.

We observe that the LDF potential is clearly the best choice among the cases studied. It will
be employed for our further calculations in this work. The residual electron correlation can be
estimated by taking the difference of the KS and LDF results. Table~\ref{tab:Li} shows that such
difference yields a conservative estimate in the case of $2p_J$ states, whereas for the $2s$ state
it is only just consistent with the deviation of the LDF value from the exact result. We found that
the choice of the parameter $x_{\alpha} = 1/2$ in the definition of the KS potential
[Eq.~(\ref{eq8})] yields a more conservative estimate for the $2s$ state. The residual electron
correlation was thus estimated as a difference of the LDF and KS results with the parameter
$x_{\alpha} = 1/2$ for the $2s$ state and $x_{\alpha} = 2/3$ for the $2p_J$ states. The estimate
obtained in this way decreases roughly as $Z^{-2}$ with increase of $Z$ and becomes entirely
negligible ({\em i.e.}, less than $5\times10^{-6}$~a.u.) for $Z>21$ for $2s$ state and $Z>30$ for
$2p_J$ states. It can be mentioned that a small potential dependence of the MBPT results appears
again for large values of $Z$ and is due to an incomplete treatment of the negative-energy part of
the Dirac spectrum.

%%%%%%%%%%%%%%%%%%%%%%%%%%%%%%%%%%%%%%%%%%%%%%%%%%%%%%%%%%%%%%%%%%%%%%%%
%%%%%
%%%%%
%%%%%%%%%%%%%%%%%%%%%%%%%%%%%%%%%%%%%%%%%%%%%%%%%%%%%%%%%%%%%%%%%%%%%%%
\begin{figure}[t]
\centerline{
\resizebox{\columnwidth}{!}{%
  \includegraphics{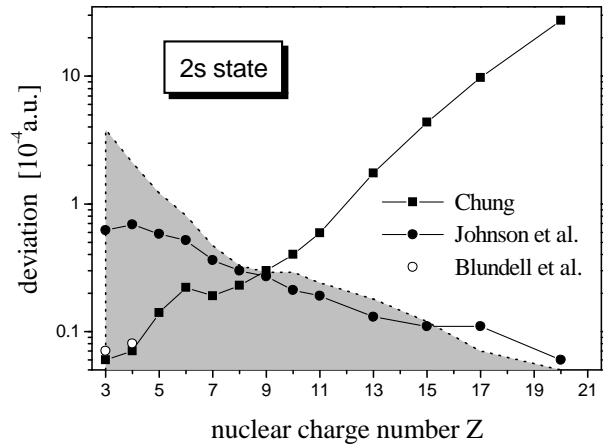}
}}
 \caption{The deviation of other calculational results from our MBPT values for the ground state
 of light Li-like ions. Squares, closed dots, and open dots indicate results by Chung
 \cite{chung:91,chung:92}, Johnson {\it et al.} \cite{johnson:88:b}, and Blundell {\em et
 al.} \cite{blundell:89:pra}, respectively. The gray filled area denotes the
 estimation of the residual electron correlation in our calculation (see text).
 \label{fig:mbpt2s}}
\end{figure}

%%%%%%%%%%%%%%%%%%%%%%%%%%%%%%%%%%%%%%%%%%%%%%%%%%%%%%%%%%%%%%%%%%%%%%%%
%%%%%
%%%%%
%%%%%%%%%%%%%%%%%%%%%%%%%%%%%%%%%%%%%%%%%%%%%%%%%%%%%%%%%%%%%%%%%%%%%%%
\begin{figure}[t]
\centerline{
\resizebox{\columnwidth}{!}{%
  \includegraphics{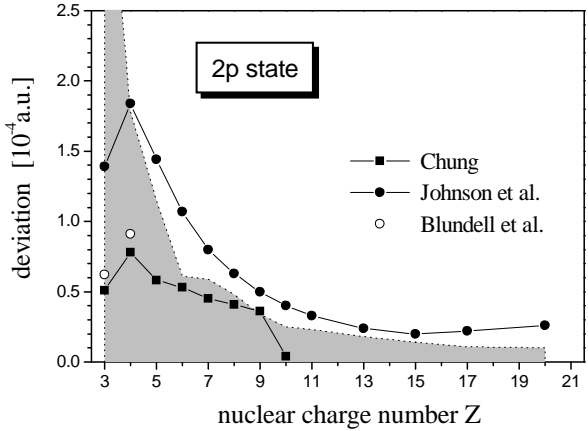}
}}
 \caption{The same as in Fig.~\ref{fig:mbpt2s}, but for the center of gravity of the
energies of the $(1s)^22p_{1/2}$ and $(1s)^22p_{3/2}$ states. The results by Chung
are from \cite{chung:93}.
 \label{fig:mbpt2p}}
\end{figure}

It is of interest to compare results of different calculations for lithium isoelectronic sequence
in the low-$Z$ region, where QED effects are small and do not significantly influence the
comparison. In Figs.~\ref{fig:mbpt2s} and \ref{fig:mbpt2p}, we plot the absolute values of the
difference of our MBPT values and the results of other calculations
\cite{johnson:88:b,blundell:89:pra,chung:91,chung:92,chung:93} for the ionization energy of the
$2s$ state and for the centroid of the $2p_{1/2}$ and $2p_{3/2}$ levels, respectively. Calculations
by Chung \cite{chung:91,chung:92,chung:93} were carried out on nonrelativistic wave functions,
first solving the Schr\"odinger equation and then adding the expectation value of the Breit-Pauli
Hamiltonian. Relativistic calculations by Blundell {\em et al.} \cite{blundell:89:pra} were
performed within a combination of the coupled-cluster formalism and MBPT. They yielded rather
accurate (by standards of MBPT) results but were carried out for lithium and beryllium only. The
evaluation by Johnson {\em et al.} \cite{johnson:88:b} is a third-order MBPT calculation that
stands most close to our treatment presented so far. In order to make a comparison with the results
of other calculations, we subtracted the contribution of the mass-polarization operator $H_4$ from
the relativistic results of Refs.~\cite{chung:91,chung:92,chung:93} and the nuclear recoil
correction from those of Refs.~\cite{blundell:89:pra,johnson:88:b}.

We would like now to outline several differences between our present MBPT treatment and the one of
Ref.~\cite{johnson:88:b}. First, we use a different starting point for the perturbative expansion
(the LDF potential instead of the nonlocal DF potential). Second, we include the Breit part of the
third-order energy correction. This is an important issue since, as previously noted in
Ref.~\cite{zherebtsov:00}, the third-order Coulomb contribution happens to be anomalously small, so
that the corresponding Breit correction is comparable with the Coulomb one even for low-$Z$ systems
and becomes dominant in the high-$Z$ region. Third, we include the negative-energy contribution for
the corrections to the wave functions by summing over the complete Dirac specturm in 
Eqs.~(\ref{eqII2}) and (\ref{eqII3}). Fourth, we employ a much larger basis set than the one
consisting of 20 positive-energy functions reported in Ref.~\cite{johnson:88:b}. This becomes
important for small values of $Z$, where significant numerical cancellations occur in our
calculations of third-order energies.

For the $2s$ state, we observe good agreement with Chung's values for the lowest-$Z$ ions and a
rapidly increasing deviation for higher-$Z$ ions. The discrepancy is mainly due to his neglect of
terms beyond the relative order $(\Za)^2$ in the one-electron energies and in the one-photon
exchange correction, which are included to all orders in relativistic calculations. Indeed, the
leading correction beyond the Breit approximation can be written within the $1/Z$ expansion as
\begin{equation} \label{bebr}
    \delta E(Z) = (\Za)^6\,a_{60}+ \alpha(\Za)^5\,a_{51}+ O[\alpha^2(\Za)^3]\,,
\end{equation}
where the coefficient $a_{60}$ originates from the expansion of the Dirac energy, $a_{60}(2s) =
-21/1024$, and $a_{51}$ comes from the one-photon exchange correction, $a_{51}(2s) = 0.1175$
\cite{safronova:98}. Resulting contribution for, {\em e.g.}, $Z=20$ is $\delta E(20) =
-0.00266$~a.u., which can be compared with the actual deviation of our result from the Chung's one
of $-0.00275$~a.u. The corresponding correction to the centroid of the $2p$ levels is much smaller
(for $Z=10$, it is about $1\times10^{-6}$~a.u.), which explains much better agreement with Chung's
results observed in Fig.~\ref{fig:mbpt2p}.

The comparison with other calculations for $Z>20$ is postponed to the next section
where we supplement our MBPT values with a rigorous QED calculation of the
two-photon exchange correction.

%%%%%%%%%%%%%%%%%%%%%%%%%%%%%%%%%%%%%%%%%%%%%%%%%%%%%%%%%%%%%%%%%%%%%%%%%%%%%%%%%%%%%%%%%%%%%%
%
%%%%%%%%%%%%%%%%%%%%%%%%%%%%%%%%%%%%%%%%%%%%%%%%%%%%%%%%%%%%%%%%%%%%%%%%%%%%%%%%%%%%%%%%%%%%%%
\section{QED treatment}
\label{sec:qed}

In this section we present an {\em ab initio} QED treatment of the electron correlation up to
second order of perturbation theory. Considering the first- and second-order corrections obtained
in the previous section and given by Eqs.~(\ref{eqII23}) and (\ref{eqII23a}), we observe that the
only part that involves some approximations is the two-photon exchange correction $\Delta
E^{(2,0|2)}$. Calculations of this correction for Li-like ions were previously performed in
Refs.~\cite{yerokhin:00:prl,yerokhin:01:2ph,sapirstein:01:lamb,andreev:01,artemyev:03}. 
Our present task will
be to carry out a similar calculation for the same model potentials as in our MBPT treatment.

General formulas for the two-photon exchange correction are derived by the two-time Green's
function method \cite{shabaev:90:ivf,shabaev:02:rep}. The correction can be conveniently
represented by a sum of the two-electron (``2el'') and three-electron (``3el'') contributions, each
of which is subdivided into the irreducible (``ir'') and the reducible (``red'') parts,
\begin{equation}  \label{2el0}
 \Delta E^{(2,0|2)} = \Delta E^{\rm 2el}_{\rm ir}+ \Delta E^{\rm 2el}_{\rm red}
   + \Delta E^{\rm 3el}_{\rm ir} + \Delta E^{\rm 3el}_{\rm red}\,.
\end{equation}
The irreducible two-electron part is given by
\begin{eqnarray} \label{2el1}
\Delta E_{\rm ir}^{\rm 2el} &=& \sum_{n_1 n_2}{}^{^{\prime}}
     \frac{i}{2\pi} \intinf d\omega\,
\nonumber \\ && \times
     \left[
    \frac{F_{\rm lad,dir}(\omega,n_1n_2)}
       {(\vare_c-\omega-u\vare_{n_1})
        (\vare_v+\omega-u\vare_{n_2})} \right. \nonumber \\
&& + \frac{F_{\rm lad,ex}(\omega,n_1n_2)}
       {(\vare_v-\omega-u\vare_{n_1})
        (\vare_c+\omega-u\vare_{n_2})} \nonumber \\
&& + \frac{F_{\rm cr,dir}(\omega,n_1n_2)}
       {(\vare_c-\omega-u\vare_{n_1})
        (\vare_v-\omega-u\vare_{n_2})} \nonumber \\
&& +\left.  \frac{F_{\rm cr,ex}(\omega,n_1n_2)}
       {(\vare_v-\omega-u\vare_{n_1})
        (\vare_v-\omega-u\vare_{n_2})} \right] \ ,
\end{eqnarray}
where
\begin{eqnarray} \label{2el2}
F_{\rm lad,dir}(\omega,n_1n_2) &=& \sum_{\mu_c \mu_1 \mu_2}
    I_{{c}{v}{n_1}{n_2}}(\omega)\,
    I_{{n_1}{n_2}{c}{v}}(\omega)\, , \\
    \label{2el2a}
F_{\rm lad,ex}(\omega,n_1n_2) &=& -\sum_{\mu_c \mu_1 \mu_2}
    I_{{v}{c}{n_1}{n_2}}(\omega)\,
    I_{{n_1}{n_2}{c}{v}}(\omega-\Delta)\, , \nonumber
    \\
    \label{2el2b}
    \\
%\end{eqnarray}
%\begin{eqnarray}
F_{\rm cr,dir}(\omega,n_1n_2) &=& \sum_{\mu_c \mu_1 \mu_2}
    I_{{c}{n_2}{n_1}{v}}(\omega)\,
    I_{{n_1}{v}{c}{n_2}}(\omega)\, , \\
    \label{2el2c}
F_{\rm cr,ex}(\omega,n_1n_2) &=& -\sum_{\mu_c \mu_1 \mu_2}
    I_{{v}{n_2}{n_1}{v}}(\omega)\,
    I_{{n_1}{c}{c}{n_2}}(\omega-\Delta)\, ,
    \nonumber \\
\end{eqnarray}
$u = (1-i0)$, $\Delta = \vare_v-\vare_c$, and $\mu_1$ and $\mu_2$ denote the momentum projections
of the states $n_1$ and $n_2$, respectively. The prime on the sum indicates that some intermediate
states are excluded from the summation. Specifically, the omitted terms are:
$(\vare_{n_1}\vare_{n_2}) = (\vare_c\vare_v), (\vare_v\vare_c)$ in the first and second terms in
the brackets, $(\vare_{n_1}\vare_{n_2}) = (\vare_c\vare_v)$ in the third term, and
$(\vare_{n_1}\vare_{n_2}) = (\vare_c\vare_c), (\vare_v\vare_v)$ in the fourth term. The above
conditions for omitting terms should be taken in the point-nucleus limit, {\em i.e.}, the $2s$ and
the $2p_{1/2}$ states are treated in the same way despite the fact that their Dirac energies are
separated by the finite nuclear size.

The reducible two-electron part is
\begin{eqnarray} \label{2el8}
\Delta E_{\rm red}^{\rm 2el} &=&
    \frac{i}{4\pi} \int^{\infty}_{-\infty}
    d\omega \, \frac{1}{(\omega+i0)^2}
    \left[ 2\,F_{\rm cr,ex}(-\omega+\Delta,cc)
      \right.
\nonumber \\ &&
    + 2\,F_{\rm cr,ex}(-\omega,vv)
    + 2\,F_{\rm cr,ex}(-\omega+\Delta_{vs},ss)
\nonumber \\ &&
     - F_{\rm lad,ex}(\omega+\Delta,cv)
     - F_{\rm lad,ex}(-\omega+\Delta,cv)
\nonumber \\ &&
    -F_{\rm lad,dir}(\omega-\Delta,vc)
    -F_{\rm lad,dir}(-\omega-\Delta,vc)
\nonumber \\ && \left.
     - F_{\rm lad,ex}(\omega,vc)
     - F_{\rm lad,ex}(-\omega,vc)
 \right] \, ,
\end{eqnarray}
where $s$ denotes the Dirac state whose energy is
separated from the valence energy $\vare_v$ by the finite
nuclear size only. (The corresponding contribution should be omitted if there is no
such state, {\it e.g.}, for $v=2p_{3/2}$.)

The irreducible three-electron contribution reads
\begin{eqnarray} \label{thr9}
\Delta E^{\rm 3el}_{\rm ir} &=& \sum_{PQ} (-1)^{P+Q}
     \sum_{n}{}^{^{\prime}}
 \nonumber \\ && \times
 \frac{\I{P2}{P3}{n}{Q3}(\Delta_{P3Q3})\,
       \I{P1}{n}{Q1}{Q2}(\Delta_{Q1P1}) }
    {\vare_{Q1}+\vare_{Q2}-\vare_{P1}-\vare_n} \, ,
 \nonumber \\
\end{eqnarray}
where the prime on the sum indicates that terms with the vanishing denominator should be omitted.
Finally, the reducible three-electron part is given by
\begin{eqnarray} \label{thr14}
 \Delta E^{\rm 3el}_{\rm red} &=& \sum_{\mu_a\mu_{\overline{v}}}
  \Bigr[
    I\pr_{{v}{a}{a}{\overline{v}}}(\Delta)\,
       \left( I_{{a}{b;}{a}{b}} - I_{{b}{v;}{b}{v}} \right)
\nonumber \\ &&
    + \frac12\, I\pr_{{a}{v}{\overline{v}}{b}}(\Delta)\, I_{{b}{\overline{v};}{a}{v}}
    + \frac12\, I\pr_{{b}{\overline{v}}{v}{a}}(\Delta)\, I_{{v}{a;}{\overline{v}}{b}}
    \Bigl] \, ,
 \nonumber \\
\end{eqnarray}
where we used the notation $I_{{a}{b;}{c}{d}}= I_{{a}{b}{c}{d}}(\Delta_{bd}) -
I_{{b}{a}{c}{d}}(\Delta_{ad})$, $a$ and $b$ denote the core electrons with the
momentum projections $\mu_a$ and $\mu_b$, $\mu_b = -\mu_a$, and $\overline{v}$ is a
valence state with the momentum projection $\mu_{\overline{v}}$.

We mention that the formulas (\ref{2el1})-(\ref{thr14}) reduce to the MBPT expression (\ref{eqII5})
after neglecting (i) the energy dependence of the operator $I(\omega)$ in the Coulomb gauge and
(ii) the negative-energy part of the Dirac spectrum. Within this approximation, all reducible parts
vanish and the $\omega$ integration can be performed by Cauchy's theorem.

%%%%%%%%%%%%%%%%%%%%%%%%%%%%%%%%%%%%%%%%%%%%%%%%%%%%%%%%%%%%%%%%%%%%%%%%%%%
%
%       Z=30
%
%%%%%%%%%%%%%%%%%%
\begin{table}[t]
\caption{Electronic-structure contributions to the ionization potential of $n=2$ states of Li-like
Zn ($Z=30$) for different model potentials, in a.u.
 \label{tab:Zn} }
\begin{ruledtabular}
\begin{tabular}{l...}
                &   \multicolumn{1}{c}{  Coul  }
                    &   \multicolumn{1}{c}{ KS}            &  \multicolumn{1}{c}{LDF} \\
\hline\\[-9pt]
 $2s_{1/2}$ state: \\
 Dirac:          &   -114.2x2811   &   -102.2x6718   &   -102.2x5381   \\
 1-photon:       &     12.2x3022   &      0.0x0568   &     -0.0x0614   \\
 2-photon(MBPT): &     -0.2x6803   &     -0.0x0493   &     -0.0x0644   \\
 2-photon(QED):  &     -0.0x0012   &     -0.0x0010   &     -0.0x0010   \\
 3-photon(MBPT): &     -0.0x0039   &      0.0x0007   &      0.0x0004   \\
  Sum:           &   -102.2x6643   &   -102.2x6647   &   -102.2x6646   \\
 [5pt]
 $2p_{1/2}$ state: \\
 Dirac:          &   -114.2x2864   &    -99.8x5790   &   -100.1x7793   \\
 1-photon:       &     14.5x1407   &     -0.2x6002   &      0.0x5156   \\
 2-photon(MBPT): &     -0.4x1102   &     -0.0x0935   &     -0.0x0066   \\
 2-photon(QED):  &     -0.0x0007   &     -0.0x0006   &     -0.0x0006   \\
 3-photon(MBPT): &     -0.0x0141   &      0.0x0016   &     -0.0x0007   \\
  Sum:           &   -100.1x2706   &   -100.1x2717   &   -100.1x2717   \\
 [5pt]
 $2p_{3/2}$ state: \\
 Dirac:          &   -112.8x3902   &    -98.7x3159   &    -99.0x5651   \\
 1-photon:       &     14.1x8655   &     -0.3x0055   &      0.0x2088   \\
 2-photon(MBPT): &     -0.3x8556   &     -0.0x0819   &     -0.0x0462   \\
 2-photon(QED):  &      0.0x0010   &      0.0x0009   &      0.0x0009   \\
 3-photon(MBPT): &     -0.0x0209   &      0.0x0011   &      0.0x0003   \\
  Sum:           &    -99.0x4002   &    -99.0x4013   &    -99.0x4014   \\
\end{tabular}
\end{ruledtabular}
\end{table}

%%%%%%%%%%%%%%%%%%%%%%%%%%%%%%%%%%%%%%%%%%%%%%%%%%%%%%%%%%%%%%%%%%%%%%%%%%%
%
%       Z=83
%
%%%%%%%%%%%%%%%%%%
\begin{table}[t]
\caption{Electronic-structure contributions to the ionization potential of $n=2$ states of Li-like
Bi ($Z=83$) for different model potentials, in a.u.
 \label{tab:Bi} }
\begin{ruledtabular}
\begin{tabular}{l...}
                &   \multicolumn{1}{c}{  Coul  }
                    &   \multicolumn{1}{c}{ KS}            &  \multicolumn{1}{c}{LDF} \\
\hline\\[-9pt]
 $2s_{1/2}$ state: \\
 Dirac:          &    -984.x4415   &    -944.x2155   &    -944.x5857   \\
 1-photon:       &      40.x8495   &       0.x2180   &       0.x5892   \\
 2-photon(MBPT): &      -0.x4253   &      -0.x0185   &      -0.x0195   \\
 2-photon(QED):  &      -0.x0010   &      -0.x0010   &      -0.x0009   \\
 3-photon(MBPT): &       0.x0015   &       0.x0002   &       0.x0001   \\
  Sum:           &    -944.x0167   &    -944.x0167   &    -944.x0168   \\
 Ref.~\cite{sapirstein:01:lamb} &   -944.x0162    &   -944.x0155    & \\
 [5pt]
 $2p_{1/2}$ state: \\
 Dirac:          &    -984.x8788   &    -934.x8717   &    -936.x0608   \\
 1-photon:       &      51.x3848   &       0.x6190   &       1.x7683   \\
 2-photon(MBPT): &      -0.x8106   &      -0.x0473   &      -0.x0069   \\
 2-photon(QED):  &       0.x0004   &       0.x0003   &       0.x0004   \\
 3-photon(MBPT): &       0.x0053   &       0.x0005   &      -0.x0002   \\
  Sum:           &    -934.x2989   &    -934.x2992   &    -934.x2993   \\
 Ref.~\cite{sapirstein:01:lamb} &   -934.x3019    &   -934.x2995    & \\
 [5pt]
 $2p_{3/2}$ state: \\
 Dirac:          &    -881.x8298   &    -839.x8108   &    -841.x1109   \\
 1-photon:       &      41.x7248   &      -0.x7621   &       0.x5326   \\
 2-photon(MBPT): &      -0.x4958   &      -0.x0270   &      -0.x0216   \\
 2-photon(QED):  &       0.x0106   &       0.x0102   &       0.x0102   \\
 3-photon(MBPT): &       0.x0008   &       0.x0000   &      -0.x0001   \\
  Sum:           &    -840.x5893   &    -840.x5898   &    -840.x5898   \\
Ref.~\cite{sapirstein:01:lamb}  &   -840.x5902    &   -840.x5896    &    \\
\end{tabular}
\end{ruledtabular}
\end{table}

The numerical evaluation of expressions (\ref{2el1})-(\ref{thr14}) was performed by using the
scheme described in detail in Refs.~\cite{yerokhin:00:prl,yerokhin:01:2ph,artemyev:03}. Summations
over the Dirac spectrum were carried out by using the dual-kinetic-balance basis set
\cite{shabaev:04:DKB} constructed with B-splines; the typical value for the number of splines was
65. The numerical uncertainty of our results was estimated to be of about $1 \times 10^{-5}$~a.u.

The results of our calculations for the $n=2$ states of Li-like zinc ($Z=30$) and bismuth ($Z=83$)
are presented in Tables~\ref{tab:Zn} and \ref{tab:Bi}, respectively. The entry ``2-photon(QED)''
represents the difference of the two-photon exchange correction calculated within QED and within
MBPT. For the moment, we disregard uncertainties due to the nuclear size and the higher-order
correlation; the purpose of Tables~\ref{tab:Zn} and \ref{tab:Bi} is to illustrate the convergence
of results obtained for different model potentials.

We observe that the total values obtained with the KS and the LDF potential almost coincide with
each other for the both cases studied, whereas the Coulomb-potential value is slightly apart from
the other two. The third-order correction turns out to be rather small for the KS and LDF
potentials but is numerically significant in the Coulomb case. In the case of bismuth, we compare
our results with the ones obtained previously by Sapirstein and Cheng \cite{sapirstein:01:lamb},
whose approach was very similar to ours. We report a good agreement with their calculation for the
results obtained with the KS potential (a deviation for the $2s$ state is mainly due to a different
nuclear radius used in that work). However, the results obtained with the Coulomb potential exhibit
a small but noticeable difference. This is due to a certain disagreement in values for the
three-photon exchange correction. For the KS potential and $Z=83$, this correction is nearly
negligible so that the discrepancy does not play any role. It can be mentioned that the differences
between our Coulomb and KS results are much smaller than those reported by Sapirstein and Cheng.

Our Coulomb results for the three-photon exchange correction and $Z=30$ presented in
Table~\ref{tab:Zn} are in reasonable agreement with Andreev {\em et al.} \cite{andreev:01}, who
obtained $-0.00044$~a.u. for the $2s$ state and $-0.0015$~a.u. for the $2p_{1/2}$ state. However,
we disagree with them for the contribution induced by two Breit and one Coulomb interactions (which
we calculated but did not include into the total values); our result for the $2p_{1/2}$ state is
six times smaller.

%%%%%%%%%%%%%%%%%%%%%%%%%%%%%%%%%%%%%%%%%%%%%%%%%%%%%%%%%%%%%%%%%%%%%%%%
%%%%%
%%%%%
%%%%%%%%%%%%%%%%%%%%%%%%%%%%%%%%%%%%%%%%%%%%%%%%%%%%%%%%%%%%%%%%%%%%%%%
\begin{figure*}
\centerline{
\resizebox{\textwidth}{!}{%
  \includegraphics{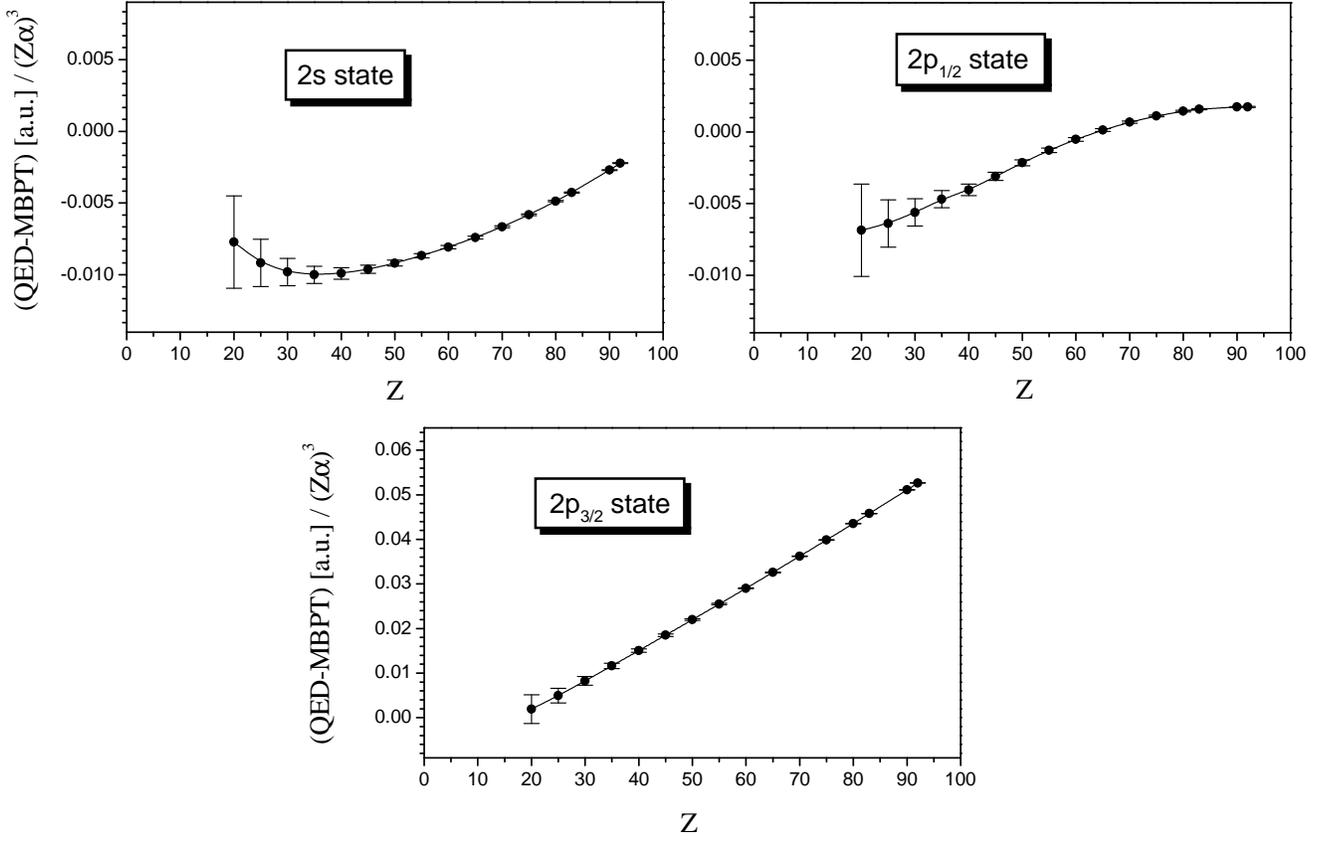}
}}
 \caption{The difference of the QED and MBPT results for the two-photon exchange
correction for the $2s$, $2p_{1/2}$, and $2p_{3/2}$ states of Li-like ions, scaled by a factor of
$(\Za)^3$.
 \label{fig:qed}}
\end{figure*}

We performed our QED calculations of the two-photon exchange correction for the nuclear charge
numbers in the range $10 \le Z\le 92$. At the lowest value considered, $Z=10$, the difference of
the QED and MBPT results is already smaller than the accuracy we are presently interested in,
$1\times 10^{-5}$~a.u., for all $n=2$ states. The difference scales as $Z^3$ and is considered
negligible for $Z<10$. In Fig.~\ref{fig:qed}, we plot this difference scaled by a factor of
$(\Za)^3$ as a function of the nuclear charge number. We observe that the ``pure'' QED part of the
two-photon exchange correction is remarkably small for the $2s$ and $2p_{1/2}$ states even in the
high-$Z$ region but is much larger in the case of the $2p_{3/2}$ state.

Our total results for the electronic-structure corrections to the ionization potential of the $n=2$
states of Li-like ions are collected in Table~\ref{tab:total}. The values for the lightest atoms
with $Z=3$ and 4 are given only for the completeness; by far more accurate calculations are
available for these systems \cite{blundell:89:pra,yan:98:prl,puchalski:06}.

The column labeled ``Dirac'' contains the energy values (minus $mc^2$) obtained from the Dirac
equation with the Fermi-like nuclear potential. The values for the the nuclear-charge
root-mean-square (rms) radii and their uncertainties are listed in the second column of the table.
They were taken from Ref.~\cite{angeli:04} except for few cases with no experimental data available
($Z=43$, 61, 85, 89, and 91), in which case we used the interpolation formula from
Ref.~\cite{johnson:85} and assigned an uncertainty of 1\% to these values. The dependence of the
Dirac value on the nuclear model was conservatively estimated by comparing the results obtained
within the Fermi and the homogeneously-charged-sphere models.

The entries ``1-photon'', ``2-ph.[MBPT]'', and ``3-photon'' contain results for the one-, two-, and
three-photon exchange corrections evaluated within MBPT with the LDF potential and given by
Eqs.~(\ref{eqII23})-(\ref{eqII24}), respectively. The entry ``2-ph.[QED]'' represents the
difference of the two-photon exchange correction evaluated within QED and MBPT, i.e., the
difference of Eqs.~(\ref{2el0}) and (\ref{eqII23a}). The numerical uncertainty of $1\times 
10^{-5}$~a.u. is not specified explicitly in Table~\ref{tab:total} but included into the total
error estimate.

The column ``h.o.'' contains errors due the higher-order effects neglected in the present
investigation. Our estimations for these effects consist of two parts that are added quadratically,
the residual electron correlation and the QED part of the three-photon exchange correction. The
residual correlation was discussed in the previous section; it has its maximum for $Z=3$ and
decreases rapidly when $Z$ increases. Making a comparison with more precise calculations available
for lithium and beryllium allow us to be reasonably confident of this part of our error estimate.

An estimation of residual three-photon QED effects is more difficult to make reliably.
Fig.~\ref{fig:qed} leads us to surmise that QED part of the two-photon exchange correction is
anomalously small for the $2s$ and $2p_{1/2}$ states; moreover, it changes its sign when $Z$ is
varied (for $2p$ states). An estimate based on the ratio of the QED and MBPT two-photon
contributions would thus likely to underestimate the three-photon QED effects. We choose to base
our estimation on the three-photon MBPT contribution induced by two Breit and one Coulomb
interactions; let us denote it by $\Delta E_{\rm MBPT}^{(3)}(B\times B)$. It has the same scaling
order as the QED contribution [$\alpha^3(\Za)^2$] and it does not change its sign through the $Z$
range of interest. For the $2s$ and $2p_{1/2}$ states, we take the absolute value of this
contribution for the estimation of three-photon QED effects. For the $2p_{3/2}$ state, we multiply
its value by the ratio
$$
\left| \frac{\Delta E^{(2)}-\Delta E_{\rm MBPT}^{(2)}}{\Delta E_{\rm MBPT}^{(2)}(B\times B)}
\right|\,,
$$
whose numerical contribution varies from 1 in the middle-$Z$ region to 5 in the high-$Z$ region.

%%%%%%%%%%%%%%%%%%%%%%%%%%%%%%%%%%%%%%%%%%%%%%%%%%%%%%%%%%%%%%%%%%%%%%%%
%%%%%
%%%%%
%%%%%%%%%%%%%%%%%%%%%%%%%%%%%%%%%%%%%%%%%%%%%%%%%%%%%%%%%%%%%%%%%%%%%%%
\begin{figure*}
\centerline{
\resizebox{\textwidth}{!}{%
  \includegraphics{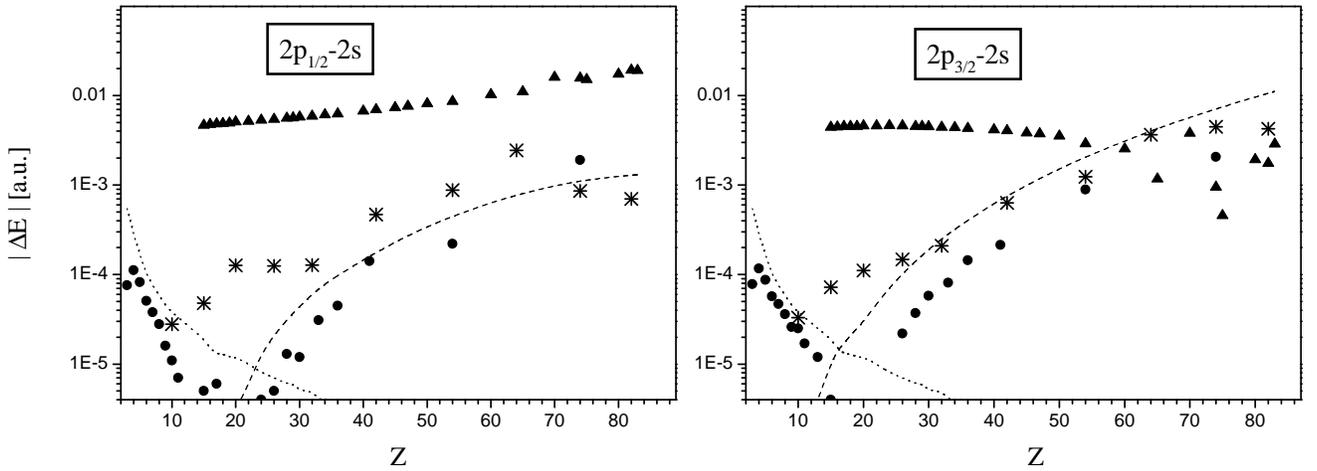}
}}
 \caption{The deviation of the present results for the the electronic-structure
part of the $2p_{1/2}$-$2s$ and $2p_{3/2}$-$2s$ transition energies from the previous calculations
(dots, MBPT \cite{johnson:88:b}; stars, CI \cite{chen:95}; triangles, MCDF \cite{indelicato:90}).
The dotted line represents the estimated error of our results due to the residual electron
correlation; the dashed line denotes the QED part of the two-photon exchange correction accounted
for in the present investigation.
 \label{fig:comparison}}
\end{figure*}

We are now in a position to compare our numerical results for the electronic-structure part of the
$2p_J$-$2s$ transition energies with results of other investigations. We selected three extensive
investigations of the electronic structure of Li-like ions to compare with, which were accomplished
by three independent methods: MBPT \cite{johnson:88:b}, the multiconfigurational Dirac-Fock (MCDF)
method \cite{indelicato:90}, and the configuration interaction (CI) method \cite{chen:95}. It
should be noted that none of these calculations accounted for the QED part of the two-photon
exchange correction, which is included into consideration in the present work.

In Fig.~\ref{fig:comparison}, we plot the deviation of our numerical values for the
electronic-structure part of the $2p_{3/2}$-$2s$ and $2p_{1/2}$-$2s$ transition energies of Li-like
ions from the results of previous calculations. The electronic-structure part can be unambiguously
isolated from the data presented in Refs.~\cite{johnson:88:b,chen:95}; from
Ref.~\cite{indelicato:90}, we had to subtract the mass-polarization term, which is explicitly given
there for $Z=15$, 54 and 92 only. In Fig.~\ref{fig:comparison}, the dotted line indicates the
estimated error of our results due to the residual correlation; the dashed line stands for the QED
part of the two-photon exchange correction, which is accounted for in our calculation but is
omitted in previous studies.

We observe a distinct and nearly $Z$-independent deviation of our results from the MCDF values,
which is on the level of about 0.005-0.01~a.u. for the $2p_{1/2}$-$2s$ transition and of about
0.003~a.u. for the $2p_{3/2}$-$2s$ transition. Agreement with the MBPT and CI results is much
better; one can observe that the main part of the deviation is, as expected, due to the QED effects
not accounted for in the previous studies.

%%%%%%%%%%%%%%%%%%%%%%%%%%%%%%%%%%%%%%%%%%%%%%%%%%%%%%%%%%%%%%%%%%%%%%%%%%%
%
%%%%%%%%%%%%%%%%%%%%%%%%%%%%%%%%%%%%%%%%%%%%%%%%%%%%%%%%%%%%%%%%%%%%%%%%%%%

\section{Transition energies in Li-like ions}
\label{sec:transitions}

%%%%%%%%%%%%%%%%%%%%%%%%%%%%%%%%%%%%%%%%%%%%%%%%%%%%%%%%%%%%%%%%%%%%%%%%%%%
%
%       2p1-2s transition
%
%%%%%%%%%%%%%%%%%%
\begin{table*}
\setlength{\LTcapwidth}{\textwidth}
\caption{Individual contributions to the $2p_{1/2}$-$2s$ transition energy in Li-like iron, nickel, and
krypton, in eV ($1~\mbox{a.u.} = 27.211\,383~\mbox{eV}$).
 \label{tab:2p1} }
\begin{ruledtabular}
\begin{tabular}{ll...}
                &  Subset & \multicolumn{1}{c}{$Z=26$}
                                         & \multicolumn{1}{c}{$Z=28$}
                                                                     &  \multicolumn{1}{c}{$Z=36$} \\
\hline\\[-9pt]
\multicolumn{2}{l}{Electronic structure}
                     &  49.103x1\,(5)     &        53.603x6\,(3)      &       72.801x3\,(6) \\
One-loop QED &       &  -0.556x5          &        -0.716x8           &       -1.685x9     \\
Screened QED &       &   0.064x9\,(20)    &         0.077x8\,(22)     &        0.143x3\,(32)\\
Recoil       &       &  -0.012x1\,(19)    &        -0.013x7\,(20)     &       -0.016x3\,(18)\\
Two-loop QED & SESE  &   0.000x4\,(1)     &         0.000x6\,(1)      &        0.002x2\,(4) \\
             & SEVP  &  -0.000x2          &        -0.000x2           &       -0.000x8     \\
             & VPVP  &   0.000x4          &         0.000x5           &        0.001x5\,(1) \\
             & S(VP)E&  -0.000x1          &        -0.000x1           &       -0.000x2\,(2) \\
\hline
Total theory &       &  48.600x0\,(28)    &        52.951x7\,(30)     &       71.245x1\,(37)\\
Experiment   &       &  48.599x7\,(10)\ \mbox{\cite{reader:94}}    
                                          &        52.950x1\,(11)\  \mbox{\cite{sugar:93}} &
                                                                              71.243x\,(8)\ \mbox{\cite{madzunkov:02}}\\
             &       &  48.603x3\,(19)\ \mbox{\cite{hinnov:89}}    
                                          &        52.949x7\,(22)\  \mbox{\cite{hinnov:89}}  &
                                                                              71.241x\,(11)\ \mbox{\cite{hinnov:89}}\\
             &       &  48.600x1\,(19)\ \mbox{\cite{knize:91}}    
                                          &        52.947x\,(4)\  \mbox{\cite{staude:98}} &  \\
\end{tabular}
\end{ruledtabular}
\end{table*}

%%%%%%%%%%%%%%%%%%%%%%%%%%%%%%%%%%%%%%%%%%%%%%%%%%%%%%%%%%%%%%%%%%%%%%%%%%%
%
%       2p3-2s transition
%
%%%%%%%%%%%%%%%%%%
\begin{table*}
\setlength{\LTcapwidth}{\textwidth}
\caption{Individual contributions to the $2p_{3/2}$-$2s$ transition energy in Li-like iron, nickel, and
silver, in eV.
 \label{tab:2p3} }
\begin{ruledtabular}
\begin{tabular}{ll...}
                & Subset  & \multicolumn{1}{c}{$Z=26$}
                               & \multicolumn{1}{c}{$Z=28$}
                                              &  \multicolumn{1}{c}{$Z=47$} \\
\hline\\[-9pt]
Electronic structure&&  65.03x33\,(5)   &   75.56x34\,(3)   &     307.19x88\,(16) \\
One-loop QED &       &  -0.51x19        &   -0.65x74        &      -3.74x39        \\
Screened QED &       &   0.05x51\,(17)  &    0.06x58\,(19)  &       0.23x16\,(39)  \\
Recoil       &       &  -0.01x23\,(20)  &   -0.01x40\,(21)  &      -0.02x43\,(19)  \\
Two-loop QED & SESE  &   0.00x03\,(1)   &    0.00x05\,(1)   &       0.00x72\,(20)  \\
             & SEVP  &  -0.00x02        &   -0.00x02        &      -0.00x32        \\
             & VPVP  &   0.00x04        &    0.00x05        &       0.00x47\,(5)   \\
             & S(VP)E&  -0.00x01        &   -0.00x01        &      -0.00x07\,(7)   \\
\hline
Total theory &       &  64.56x47\,(26)  &   74.95x85\,(28)  &     303.67x04\,(52)  \\
Experiment   &       &  64.56x56\,(17)\ \mbox{\cite{reader:94}}    
                                       &   74.96x02\,(22)\ \mbox{\cite{sugar:93}} 
                                                           &     303.67x\,(3)\ \mbox{\cite{bosselmann:99}}\\
             &       &  64.57x1\,(7)\ \mbox{\cite{hinnov:89}}    
                                       &   74.96x2\,(4)\  \mbox{\cite{hinnov:89}} & \\
             &       &  64.56x0\,(3)\ \mbox{\cite{knize:91}}    
                                       &   74.95x8\,(7)\  \mbox{\cite{staude:98}} &  \\
\end{tabular}
\end{ruledtabular}
\end{table*}

A number of important corrections should be added to the electronic-structure part of transition
energies addressed to in the previous section in order to allow an adequate comparison with
experimental data. We now briefly discuss various theoretical contributions to the $2p_{1/2}$-$2s$
and $2p_{3/2}$-$2s$ transition energies in Li-like ions. Our discussion is summarized by Tables
\ref{tab:2p1} and \ref{tab:2p3}, where various individual contributions are collected for several
medium-$Z$ ions.

The largest effect to be added to the electronic-structure part is the one-loop QED contribution.
It consists of the self-energy and vacuum-polarization corrections and is presently 
well-understood; we refer the reader to the review \cite{mohr:98} for the details.
The numerical values listed in Tables~\ref{tab:2p1} and \ref{tab:2p3} represent the one-loop QED
correction calculated on the hydrogenic wave functions.

The next effect is the screening of the one-loop QED corrections by other electrons. Rigorous
evaluations of the first-order (in $1/Z$) screening effects were performed for Li-like ions in our
previous investigations \cite{yerokhin:99:sescr,artemyev:99,yerokhin:05:OS}; a similar calculation
was carried out for Li-like bismuth by Sapirstein and Cheng \cite{sapirstein:01:lamb}. Higher-order
screening effects have not been calculated up to now. The error due to their neglect has to be
estimated with some care since it presently yields the dominant theoretical uncertainty for
medium-$Z$ ions. To obtain such an estimate, we recall that, within the $\Za$ expansion, the
dominant part of the screening of one-electron QED corrections can be described
\cite{araki:57,sucher:58} by incorporating the correct electron density at the nucleus into the
hydrogenic formulas. For the $2p$-$2s$ transition, the nonrelativistic electron density at the
origin is given by \cite{mckenzie:91}
\begin{align} \label{delta}
 & \left< \textstyle \sum_i \delta(\bfr_i)\right>_{(1s)^2 2p} -
     \left< \textstyle \sum_i \delta(\bfr_i)\right>_{(1s)^2 2s}
     \nonumber \\
  & = \frac{(\Za)^3}{\pi} \left[ -1/8 + 0.339\,893\,Z^{-1}- 0.139\,Z^{-2} + O(Z^{-3})\right]\,.
\end{align}
The ratio of the second and the first term in the above expression ($\approx-2.7/Z$) is very close
to the actual ratio of the first-order screening contribution to the hydrogenic one. We estimate
the ratio of the higher-order effects to the first-order screening by taking the ratio of the
$Z^{-2}$ term in Eq.~(\ref{delta}) to the $Z^{-1}$ one and multiplying it by a conservative factor
of 2, the resulting scaling factor thus being $0.8/Z$.

Another important effect to be taken into account is the recoil correction. Rigorous QED
calculations of the leading term of the $1/Z$ expansion of this correction were performed in
Refs.~\cite{artemyev:95:pra,artemyev:95:jpb}. The higher-order (in $1/Z$) part of the recoil effect
has been addressed to in the case of Li-like ions only nonrelativistically up to now. 
Within the nonrelativistic approximation,
we obtain it by evaluating the reduced-mass correction to the electronic-structure part of order
$1/Z$ and higher and (for the $2p$ states) by adding the part of the mass polarization of order
$1/Z$ and higher inferred from the Hughes-Eckart formula \cite{hughes:30}. A 100\% 
uncertainty was acribed to the part of the recoil effect obtained within the nonrelativistic 
approximation. 

Finally, we should account for the two-loop QED effects. The first complete all-order calculation of
the two-loop one-electron QED corrections for $n=2$ states was recently accomplished in
Ref.~\cite{yerokhin:06:prl} for several ions with $Z\ge 60$. In order to complete our compilation
for ions with $Z=26$, 28, 36, and 47 in Tables \ref{tab:2p1} and \ref{tab:2p3}, we need values for the
two-loop QED corrections outside the $Z$ range covered in Ref.~\cite{yerokhin:06:prl}. To this end,
we performed calculations of the two-loop diagrams involving closed electron loops. In notations of
Ref.~\cite{yerokhin:06:prl}, these are the subsets ``SEVP'', ``VPVP'', and ``S(VP)E''. The results
are presented in Tables \ref{tab:2p1} and \ref{tab:2p3}; the error bars indicated are due to the
free-loop approximation employed in the evaluation of the VPVP and S(VP)E subsets, see
Ref.~\cite{yerokhin:06:prl} for details.

The remaining two-loop self-energy correction (the ``SESE'' subset) is much more difficult to
calculate and we obtain its numerical values by an extrapolation. For the $2s$ state,
the extrapolation was performed in two steps. First, we obtain numerical values for the $1s$ state,
which was done by interpolating the numerical results of Ref.~\cite{yerokhin:05:sese}. Second, we
obtain results for the weighted difference of the $2s$ and $1s$ corrections, $\Delta_s = 8\,\delta
E_{2s}-\delta E_{1s}$. This was achieved by subtracting all $\Za$-expansion contributions known
(see Refs.~\cite{czarnecki:05:prl,jentschura:05:sese} and references therein) from the all-order
results of Ref.~\cite{yerokhin:06:prl} and extrapolating the higher-order remainder towards the
values of $Z$ of interest. An uncertainty of 30\% was ascribed to these results. For the $2p$
states, the correction is much smaller and, for our purposes, it is sufficient to obtain just the
boundaries for the higher-order remainder. The comparison \cite{yerokhin:06:cjp} of the all-order
numerical values with the $\Za$-expansion results suggests an estimate of $\pm 2\,
\alpha^2(\Za)^6/(8\,\pi^2)$ for the higher-order remainder.

Considering the data presented in Tables \ref{tab:2p1} and \ref{tab:2p3}, we notice a remarkably
good agreement of our total values with the experimental results. The theoretical accuracy is 
lower than the experimental one in the case of iron and nickel but significantly better for
krypton and silver. It should be mentioned that the leading theoretical uncertainty for
medium-$Z$ ions stems presently from the higher-order screening and recoil effects. In this respect, the
situation for medium-$Z$ ions is different from the one encountered in the high-$Z$ region, where
the uncertainties due to the finite nuclear size effect and due to the two-loop QED corrections
become prominent \cite{yerokhin:06:prl}.

The uncertainty due to the screening of QED effects can be reduced by calculating the one-loop and the
first-order screening QED corrections not on hydrogenic wave functions but in the presence of a
local screening potential, as it was done in the case of bismuth in Ref.~\cite{sapirstein:01:lamb}.
The uncertainty of the recoil effect can also be improved by calculating the higher-order 
(in $1/Z$) recoil correction within the leading relativistic approximation. 
This means that the theoretical accuracy can be pushed even further in the near future.

%%%%%%%%%%%%%%%%%%%%%%%%%%%%%%%%%%%%%%%%%%%%%%%%%%%%%%%%%%%%%%%%%%%%%%%%%%%
%
%%%%%%%%%%%%%%%%%%%%%%%%%%%%%%%%%%%%%%%%%%%%%%%%%%%%%%%%%%%%%%%%%%%%%%%%%%%

\section{Conclusion}
\label{sec:conclusion}

In this paper we have presented a systematic QED treatment of the electron correlation for $n=2$
states of Li-like ions. The treatment relies on the perturbative expansion, with a local model
potential included into the zeroth-order approximation. For the first two terms of the expansion,
rigorous QED calculations were performed, whereas the third-order contribution was evaluated within
the MBPT approximation. Errors due to truncation of the perturbative expansion and due to usage of
the MBPT approximation in the third-order correction were estimated.

Collecting all theoretical contributions available for the $2p_J$-$2s$ transition energies, we
observed good agreement with experimental results for medium-$Z$ ions. 
The dominant uncertainty of the theoretical
results in the medium-$Z$ region is shown to originate from the higher-order screening of QED
corrections and from the recoil effect. 
Further improvement of the theoretical accuracy is anticipated for medium-$Z$ ions. 

\section*{Acknowledgements}

V.A.Y. is indebted to A. Surzhykov for an introduction into the GRASP package. This work was
supported by the RFBR grant No.~04-02-17574. V.A.Y. acknowledges the support by the RFBR grant
No.~06-02-04007.

%%%%%%%%%%%%%%%%%%%%%%%%%%%%%%%%%%%%%%%%%%%%%%%%%%%%%
%\bibliographystyle{c:/-doc-/papers/bibtex/phaip30}
%\bibliography{c:/-doc-/papers/bibtex/hfst}

%%%%%%%%%%%%%%%%%%%%%%%%%%%%%%%%%%%%%%%%%%%%%%%%%%%%%%%%%%%%%%%%%%%%%%%%%%%
%
%       Total table
%
%%%%%%%%%%%%%%%%%%
\newpage
\setlength{\LTcapwidth}{\textwidth}
\begingroup
\begin{ruledtabular}
\begin{longtable*}{l.c.......}
\caption{Electronic-structure contributions to the ionization potential of the $n=2$ states of
Li-like ions, in a.u.
 \label{tab:total} }
\\
\hline \hline
    $Z$         &   \multicolumn{1}{c}{ $\rms$ }
                  &  State
                    &   \multicolumn{1}{c}{Dirac}
                         &  \multicolumn{1}{c}{1-photon}
                               &  \multicolumn{1}{c}{2ph.[MBPT]}
                                     &  \multicolumn{1}{c}{2ph.[QED]}
                                               &  \multicolumn{1}{c}{3-photon}
                                                     &  \multicolumn{1}{c}{h.o.}
                                                              &  \multicolumn{1}{c}{Total}
\\
\colrule
\endfirsthead
\caption{Electronic-structure contributions (continued)}\\
\hline \hline
    $Z$         &   \multicolumn{1}{c}{ $\rms$ }
                  &  State
                    &   \multicolumn{1}{c}{Dirac}
                         &  \multicolumn{1}{c}{1-photon}
                               &  \multicolumn{1}{c}{2ph.[MBPT]}
                                     &  \multicolumn{1}{c}{2ph.[QED]}
                                               &  \multicolumn{1}{c}{3-photon}
                                                     &  \multicolumn{1}{c}{h.o.}
                                                              &  \multicolumn{1}{c}{Total}
\\
\colrule
\endhead
\hline \hline
\endfoot
  3 & 2.43x1\,(28)  & $2s_{1/2}$ &     -0.x19601       &  -0.0x0031 &  -0.0x0184 &  0.0x0000 &  0.0x0001 & \pm~0.0x0038 &      -0.x19815\,(38) \\
    &               & $2p_{1/2}$ &     -0.x12867       &  -0.0x0109 &  -0.0x0045 &  0.0x0000 & -0.0x0009 & \pm~0.0x0039 &      -0.x13029\,(39) \\
    &               & $2p_{3/2}$ &     -0.x12867       &  -0.0x0108 &  -0.0x0046 &  0.0x0000 & -0.0x0009 & \pm~0.0x0039 &      -0.x13029\,(39) \\
  4 & 2.51x8\,(11)  & $2s_{1/2}$ &     -0.x66520       &  -0.0x0117 &  -0.0x0293 &  0.0x0000 &  0.0x0000 & \pm~0.0x0021 &      -0.x66930\,(21) \\
    &               & $2p_{1/2}$ &     -0.x51971       &  -0.0x0303 &  -0.0x0095 &  0.0x0000 & -0.0x0016 & \pm~0.0x0018 &      -0.x52386\,(18) \\
    &               & $2p_{3/2}$ &     -0.x51967       &  -0.0x0298 &  -0.0x0102 &  0.0x0000 & -0.0x0016 & \pm~0.0x0018 &      -0.x52383\,(18) \\
  5 & 2.40x6\,(29)  & $2s_{1/2}$ &     -1.x38848       &  -0.0x0202 &  -0.0x0352 &  0.0x0000 &  0.0x0000 & \pm~0.0x0012 &      -1.x39402\,(12) \\
    &               & $2p_{1/2}$ &     -1.x16794       &  -0.0x0424 &  -0.0x0133 &  0.0x0000 & -0.0x0015 & \pm~0.0x0011 &      -1.x17366\,(12) \\
    &               & $2p_{3/2}$ &     -1.x16775       &  -0.0x0416 &  -0.0x0147 &  0.0x0000 & -0.0x0013 & \pm~0.0x0011 &      -1.x17351\,(12) \\
  6 & 2.47x0\,(2)   & $2s_{1/2}$ &     -2.x36360       &  -0.0x0311 &  -0.0x0354 &  0.0x0000 & -0.0x0003 & \pm~0.0x0008 &      -2.x37027\,(8) \\
    &               & $2p_{1/2}$ &     -2.x06991       &  -0.0x0472 &  -0.0x0160 &  0.0x0000 & -0.0x0012 & \pm~0.0x0006 &      -2.x07636\,(6) \\
    &               & $2p_{3/2}$ &     -2.x06933       &  -0.0x0462 &  -0.0x0182 &  0.0x0000 & -0.0x0010 & \pm~0.0x0006 &      -2.x07588\,(6) \\
  7 & 2.55x8\,(7)   & $2s_{1/2}$ &     -3.x58947       &  -0.0x0416 &  -0.0x0400 &  0.0x0000 & -0.0x0001 & \pm~0.0x0005 &      -3.x59764\,(5) \\
    &               & $2p_{1/2}$ &     -3.x22370       &  -0.0x0526 &  -0.0x0176 &  0.0x0000 & -0.0x0011 & \pm~0.0x0006 &      -3.x23083\,(6) \\
    &               & $2p_{3/2}$ &     -3.x22234       &  -0.0x0516 &  -0.0x0208 &  0.0x0000 & -0.0x0008 & \pm~0.0x0006 &      -3.x22966\,(6) \\
  8 & 2.70x1\,(6)   & $2s_{1/2}$ &     -5.x06676       &  -0.0x0508 &  -0.0x0428 &  0.0x0000 &  0.0x0000 & \pm~0.0x0003 &      -5.x07613\,(3) \\
    &               & $2p_{1/2}$ &     -4.x62925       &  -0.0x0550 &  -0.0x0186 &  0.0x0000 & -0.0x0010 & \pm~0.0x0005 &      -4.x63671\,(5) \\
    &               & $2p_{3/2}$ &     -4.x62651       &  -0.0x0544 &  -0.0x0229 &  0.0x0000 & -0.0x0006 & \pm~0.0x0005 &      -4.x63430\,(5) \\
  9 & 2.89x8\,(2)   & $2s_{1/2}$ &     -6.x79562       &  -0.0x0599 &  -0.0x0432 &  0.0x0000 & -0.0x0001 & \pm~0.0x0003 &      -6.x80593\,(3) \\
    &               & $2p_{1/2}$ &     -6.x28658       &  -0.0x0539 &  -0.0x0192 &  0.0x0000 & -0.0x0008 & \pm~0.0x0003 &      -6.x29398\,(4) \\
    &               & $2p_{3/2}$ &     -6.x28163       &  -0.0x0541 &  -0.0x0247 &  0.0x0000 & -0.0x0005 & \pm~0.0x0003 &      -6.x28955\,(4) \\
 10 & 3.00x5\,(2)   & $2s_{1/2}$ &     -8.x77583       &  -0.0x0696 &  -0.0x0457 &  0.0x0000 &  0.0x0000 & \pm~0.0x0003 &      -8.x78736\,(3) \\
    &               & $2p_{1/2}$ &     -8.x19547       &  -0.0x0531 &  -0.0x0194 &  0.0x0000 & -0.0x0008 & \pm~0.0x0003 &      -8.x20281\,(3) \\
    &               & $2p_{3/2}$ &     -8.x18719       &  -0.0x0547 &  -0.0x0261 &  0.0x0000 & -0.0x0003 & \pm~0.0x0003 &      -8.x19531\,(3) \\
 11 & 2.99x4\,(2)   & $2s_{1/2}$ &    -11.x00800       &  -0.0x0813 &  -0.0x0465 &  0.0x0000 &  0.0x0000 & \pm~0.0x0002 &     -11.x02078\,(3) \\
    &               & $2p_{1/2}$ &    -10.x35648       &  -0.0x0496 &  -0.0x0194 & -0.0x0001 & -0.0x0007 & \pm~0.0x0002 &     -10.x36347\,(3) \\
    &               & $2p_{3/2}$ &    -10.x34345       &  -0.0x0532 &  -0.0x0274 &  0.0x0000 & -0.0x0003 & \pm~0.0x0002 &     -10.x35154\,(3) \\
 12 & 3.05x7\,(2)   & $2s_{1/2}$ &    -13.x49330       &  -0.0x0863 &  -0.0x0470 &  0.0x0000 &  0.0x0000 & \pm~0.0x0002 &     -13.x50663\,(2) \\
    &               & $2p_{1/2}$ &    -12.x76996       &  -0.0x0437 &  -0.0x0192 & -0.0x0001 & -0.0x0007 & \pm~0.0x0002 &     -12.x77633\,(2) \\
    &               & $2p_{3/2}$ &    -12.x75036       &  -0.0x0500 &  -0.0x0286 &  0.0x0000 & -0.0x0002 & \pm~0.0x0002 &     -12.x75825\,(2) \\
 13 & 3.06x1\,(4)   & $2s_{1/2}$ &    -16.x23067       &  -0.0x0999 &  -0.0x0476 & -0.0x0001 &  0.0x0001 & \pm~0.0x0002 &     -16.x24542\,(2) \\
    &               & $2p_{1/2}$ &    -15.x43621       &  -0.0x0367 &  -0.0x0189 & -0.0x0001 & -0.0x0007 & \pm~0.0x0002 &     -15.x44184\,(2) \\
    &               & $2p_{3/2}$ &    -15.x40784       &  -0.0x0466 &  -0.0x0297 &  0.0x0000 & -0.0x0001 & \pm~0.0x0002 &     -15.x41549\,(2) \\
 14 & 3.12x2\,(2)   & $2s_{1/2}$ &    -19.x22244       &  -0.0x1040 &  -0.0x0488 & -0.0x0001 &  0.0x0001 & \pm~0.0x0001 &     -19.x23772\,(2) \\
    &               & $2p_{1/2}$ &    -18.x35582       &  -0.0x0277 &  -0.0x0184 & -0.0x0001 & -0.0x0007 & \pm~0.0x0002 &     -18.x36050\,(2) \\
    &               & $2p_{3/2}$ &    -18.x31603       &  -0.0x0422 &  -0.0x0307 &  0.0x0000 & -0.0x0001 & \pm~0.0x0002 &     -18.x32333\,(2) \\
 15 & 3.18x9\,(2)   & $2s_{1/2}$ &    -22.x46840       &  -0.0x1073 &  -0.0x0500 & -0.0x0001 &  0.0x0001 & \pm~0.0x0001 &     -22.x48413\,(2) \\
    &               & $2p_{1/2}$ &    -21.x52936       &  -0.0x0166 &  -0.0x0178 & -0.0x0001 & -0.0x0006 & \pm~0.0x0001 &     -21.x53287\,(2) \\
    &               & $2p_{3/2}$ &    -21.x47501       &  -0.0x0367 &  -0.0x0317 &  0.0x0000 &  0.0x0000 & \pm~0.0x0001 &     -21.x48185\,(2) \\
 16 & 3.26x1\,(2)   & $2s_{1/2}$ &    -25.x96925       &  -0.0x1096 &  -0.0x0511 & -0.0x0001 &  0.0x0001 & \pm~0.0x0001 &     -25.x98532\,(2) \\
    &               & $2p_{1/2}$ &    -24.x95748       &  -0.0x0031 &  -0.0x0172 & -0.0x0001 & -0.0x0006 & \pm~0.0x0001 &     -24.x95959\,(1) \\
    &               & $2p_{3/2}$ &    -24.x88489       &  -0.0x0301 &  -0.0x0326 &  0.0x0000 &  0.0x0000 & \pm~0.0x0001 &     -24.x89115\,(1) \\
 17 & 3.36x5\,(15)  & $2s_{1/2}$ &    -29.x72572       &  -0.0x1110 &  -0.0x0521 & -0.0x0002 &  0.0x0001 & \pm~0.0x0001 &     -29.x74203\,(1) \\
    &               & $2p_{1/2}$ &    -28.x64087       &   0.0x0126 &  -0.0x0165 & -0.0x0001 & -0.0x0006 & \pm~0.0x0001 &     -28.x64133\,(1) \\
    &               & $2p_{3/2}$ &    -28.x54578       &  -0.0x0224 &  -0.0x0335 &  0.0x0000 &  0.0x0001 & \pm~0.0x0001 &     -28.x55136\,(1) \\
 18 & 3.42x7\,(2)   & $2s_{1/2}$ &    -33.x73717       &  -0.0x1263 &  -0.0x0523 & -0.0x0002 &  0.0x0001 & \pm~0.0x0001 &     -33.x75504\,(1) \\
    &               & $2p_{1/2}$ &    -32.x58029       &   0.0x0310 &  -0.0x0157 & -0.0x0001 & -0.0x0006 & \pm~0.0x0001 &     -32.x57884\,(1) \\
    &               & $2p_{3/2}$ &    -32.x45782       &  -0.0x0135 &  -0.0x0344 &  0.0x0000 &  0.0x0001 & \pm~0.0x0001 &     -32.x46260\,(1) \\
 19 & 3.43x5\,(2)   & $2s_{1/2}$ &    -38.x00721       &  -0.0x1264 &  -0.0x0533 & -0.0x0002 &  0.0x0002 & \pm~0.0x0001 &     -38.x02519\,(1) \\
    &               & $2p_{1/2}$ &    -36.x77654       &   0.0x0521 &  -0.0x0150 & -0.0x0002 & -0.0x0006 & \pm~0.0x0001 &     -36.x77291\,(1) \\
    &               & $2p_{3/2}$ &    -36.x62116       &  -0.0x0033 &  -0.0x0353 &  0.0x0000 &  0.0x0001 & \pm~0.0x0001 &     -36.x62500\,(1) \\
 20 & 3.47x6\,(1)   & $2s_{1/2}$ &    -42.x53541       &  -0.0x1252 &  -0.0x0544 & -0.0x0002 &  0.0x0002 & \pm~0.0x0001 &     -42.x55337\,(1) \\
    &               & $2p_{1/2}$ &    -41.x23051       &   0.0x0762 &  -0.0x0142 & -0.0x0002 & -0.0x0006 & \pm~0.0x0001 &     -41.x22439\,(1) \\
    &               & $2p_{3/2}$ &    -41.x03595       &   0.0x0082 &  -0.0x0362 &  0.0x0001 &  0.0x0001 & \pm~0.0x0001 &     -41.x03873\,(1) \\
 21 & 3.54x4\,(2)   & $2s_{1/2}$ &    -47.x32269       &  -0.0x1227 &  -0.0x0557 & -0.0x0003 &  0.0x0002 & \pm~0.0x0001 &     -47.x34053\,(1) \\
    &               & $2p_{1/2}$ &    -45.x94308       &   0.0x1031 &  -0.0x0134 & -0.0x0002 & -0.0x0006 & \pm~0.0x0001 &     -45.x93420\,(1) \\
    &               & $2p_{3/2}$ &    -45.x70234       &   0.0x0209 &  -0.0x0371 &  0.0x0001 &  0.0x0002 & \pm~0.0x0001 &     -45.x70394\,(1) \\
 22 & 3.59x1\,(2)   & $2s_{1/2}$ &    -52.x37013       &  -0.0x1189 &  -0.0x0567 & -0.0x0003 &  0.0x0002 & \pm~0.0x0001 &     -52.x38770\,(1) \\
    &               & $2p_{1/2}$ &    -50.x91531       &   0.0x1335 &  -0.0x0126 & -0.0x0003 & -0.0x0006 & \pm~0.0x0001 &     -50.x90331\,(1) \\
    &               & $2p_{3/2}$ &    -50.x62056       &   0.0x0352 &  -0.0x0380 &  0.0x0001 &  0.0x0002 & \pm~0.0x0001 &     -50.x62081\,(1) \\
 23 & 3.59x9\,(2)   & $2s_{1/2}$ &    -57.x67878       &  -0.0x1138 &  -0.0x0577 & -0.0x0004 &  0.0x0003 & \pm~0.0x0001 &     -57.x69595\,(1) \\
    &               & $2p_{1/2}$ &    -56.x14820       &   0.0x1673 &  -0.0x0118 & -0.0x0003 & -0.0x0007 & \pm~0.0x0001 &     -56.x13275\,(1) \\
    &               & $2p_{3/2}$ &    -55.x79077       &   0.0x0510 &  -0.0x0389 &  0.0x0002 &  0.0x0002 & \pm~0.0x0001 &     -55.x78953\,(1) \\
 24 & 3.64x2\,(2)   & $2s_{1/2}$ &    -63.x24979       &  -0.0x1070 &  -0.0x0587 & -0.0x0005 &  0.0x0003 & \pm~0.0x0001 &     -63.x26638\,(1) \\
    &               & $2p_{1/2}$ &    -61.x64287       &   0.0x2048 &  -0.0x0110 & -0.0x0004 & -0.0x0007 & \pm~0.0x0001 &     -61.x62359\,(1) \\
    &               & $2p_{3/2}$ &    -61.x21319       &   0.0x0684 &  -0.0x0399 &  0.0x0002 &  0.0x0002 & \pm~0.0x0001 &     -61.x21029\,(1) \\
 25 & 3.70x6\,(2)   & $2s_{1/2}$ &    -69.x08434       &  -0.0x0989 &  -0.0x0598 & -0.0x0006 &  0.0x0003 & \pm~0.0x0001 &     -69.x10023\,(1) \\
    &               & $2p_{1/2}$ &    -67.x40049       &   0.0x2459 &  -0.0x0102 & -0.0x0004 & -0.0x0007 & \pm~0.0x0001 &     -67.x37702\,(1) \\
    &               & $2p_{3/2}$ &    -66.x88801       &   0.0x0874 &  -0.0x0409 &  0.0x0003 &  0.0x0002 & \pm~0.0x0001 &     -66.x88330\,(1) \\
 26 & 3.73x7\,(2)   & $2s_{1/2}$ &    -75.x18369       &  -0.0x0890 &  -0.0x0609 & -0.0x0006 &  0.0x0003 & \pm~0.0x0001 &     -75.x19871\,(1) \\
    &               & $2p_{1/2}$ &    -73.x42227       &   0.0x2912 &  -0.0x0094 & -0.0x0004 & -0.0x0007 & \pm~0.0x0001 &     -73.x39421\,(1) \\
    &               & $2p_{3/2}$ &    -72.x81547       &   0.0x1081 &  -0.0x0419 &  0.0x0004 &  0.0x0002 & \pm~0.0x0001 &     -72.x80878\,(1) \\
 27 & 3.78x8\,(2)   & $2s_{1/2}$ &    -81.x54918       &  -0.0x0775 &  -0.0x0620 & -0.0x0007 &  0.0x0003 & \pm~0.0x0001 &     -81.x56317\,(1) \\
    &               & $2p_{1/2}$ &    -79.x70952       &   0.0x3405 &  -0.0x0087 & -0.0x0005 & -0.0x0007 & \pm~0.0x0001 &     -79.x67645\,(1) \\
    &               & $2p_{3/2}$ &    -78.x99579       &   0.0x1305 &  -0.0x0429 &  0.0x0005 &  0.0x0002 & \pm~0.0x0001 &     -78.x98696\,(1) \\
 28 & 3.77x5\,(1)   & $2s_{1/2}$ &    -88.x17955       &  -0.0x0917 &  -0.0x0621 & -0.0x0008 &  0.0x0003 & \pm~0.0x0001 &     -88.x19497\,(1) \\
    &               & $2p_{1/2}$ &    -86.x26358       &   0.0x3942 &  -0.0x0080 & -0.0x0005 & -0.0x0007 & \pm~0.0x0001 &     -86.x22508\,(1) \\
    &               & $2p_{3/2}$ &    -85.x42924       &   0.0x1548 &  -0.0x0440 &  0.0x0006 &  0.0x0003 & \pm~0.0x0001 &     -85.x41807\,(1) \\
 29 & 3.88x2\,(2)   & $2s_{1/2}$ &    -95.x08143       &  -0.0x0775 &  -0.0x0632 & -0.0x0009 &  0.0x0003 & \pm~0.0x0001 &     -95.x09556\,(1) \\
    &               & $2p_{1/2}$ &    -93.x08590       &   0.0x4525 &  -0.0x0073 & -0.0x0005 & -0.0x0007 & \pm~0.0x0001 &     -93.x04150\,(1) \\
    &               & $2p_{3/2}$ &    -92.x11605       &   0.0x1808 &  -0.0x0451 &  0.0x0007 &  0.0x0003 & \pm~0.0x0001 &     -92.x10238\,(1) \\
 30 & 3.92x9\,(1)   & $2s_{1/2}$ &   -102.x25381       &  -0.0x0614 &  -0.0x0644 & -0.0x0010 &  0.0x0004 & \pm~0.0x0001 &    -102.x26646\,(1) \\
    &               & $2p_{1/2}$ &   -100.x17793       &   0.0x5156 &  -0.0x0066 & -0.0x0006 & -0.0x0007 & \pm~0.0x0001 &    -100.x12717\,(1) \\
    &               & $2p_{3/2}$ &    -99.x05651       &   0.0x2088 &  -0.0x0462 &  0.0x0009 &  0.0x0003 & \pm~0.0x0001 &     -99.x04014\,(1) \\
 31 & 3.99x7\,(2)   & $2s_{1/2}$ &   -109.x69828       &  -0.0x0434 &  -0.0x0656 & -0.0x0011 &  0.0x0004 & \pm~0.0x0001 &    -109.x70926\,(1) \\
    &               & $2p_{1/2}$ &   -107.x54126       &   0.0x5835 &  -0.0x0060 & -0.0x0006 & -0.0x0007 & \pm~0.0x0001 &    -107.x48365\,(1) \\
    &               & $2p_{3/2}$ &   -106.x25089       &   0.0x2388 &  -0.0x0474 &  0.0x0010 &  0.0x0003 & \pm~0.0x0001 &    -106.x23162\,(2) \\
 32 & 4.07x4\,(1)   & $2s_{1/2}$ &   -117.x41648       &  -0.0x0234 &  -0.0x0668 & -0.0x0013 &  0.0x0004 & \pm~0.0x0001 &    -117.x42559\,(1) \\
    &               & $2p_{1/2}$ &   -115.x17749       &   0.0x6565 &  -0.0x0055 & -0.0x0007 & -0.0x0007 & \pm~0.0x0001 &    -115.x11252\,(2) \\
    &               & $2p_{3/2}$ &   -113.x69948       &   0.0x2708 &  -0.0x0486 &  0.0x0012 &  0.0x0003 & \pm~0.0x0001 &    -113.x67710\,(2) \\
 33 & 4.09x7\,(2)   & $2s_{1/2}$ &   -125.x41020       &  -0.0x0009 &  -0.0x0681 & -0.0x0014 &  0.0x0004 & \pm~0.0x0001 &    -125.x41720\,(1) \\
    &               & $2p_{1/2}$ &   -123.x08834       &   0.0x7352 &  -0.0x0049 & -0.0x0007 & -0.0x0007 & \pm~0.0x0001 &    -123.x01546\,(2) \\
    &               & $2p_{3/2}$ &   -121.x40257       &   0.0x3049 &  -0.0x0499 &  0.0x0014 &  0.0x0003 & \pm~0.0x0001 &    -121.x37690\,(2) \\
 34 & 4.14x0\,(2)   & $2s_{1/2}$ &   -133.x68119       &   0.0x0236 &  -0.0x0694 & -0.0x0015 &  0.0x0004 & \pm~0.0x0001 &    -133.x68588\,(2) \\
    &               & $2p_{1/2}$ &   -131.x27557       &   0.0x8193 &  -0.0x0044 & -0.0x0007 & -0.0x0008 & \pm~0.0x0001 &    -131.x19424\,(2) \\
    &               & $2p_{3/2}$ &   -129.x36048       &   0.0x3411 &  -0.0x0512 &  0.0x0017 &  0.0x0003 & \pm~0.0x0002 &    -129.x33129\,(2) \\
 35 & 4.16x3\,(2)   & $2s_{1/2}$ &   -142.x23136       &   0.0x0503 &  -0.0x0707 & -0.0x0017 &  0.0x0004 & \pm~0.0x0001 &    -142.x23352\,(2) \\
    &               & $2p_{1/2}$ &   -139.x74102       &   0.0x9090 &  -0.0x0040 & -0.0x0008 & -0.0x0008 & \pm~0.0x0002 &    -139.x65067\,(2) \\
    &               & $2p_{3/2}$ &   -137.x57354       &   0.0x3795 &  -0.0x0525 &  0.0x0019 &  0.0x0003 & \pm~0.0x0002 &    -137.x54062\,(2) \\
 36 & 4.18x8\,(1)   & $2s_{1/2}$ &   -151.x06266       &   0.0x0794 &  -0.0x0720 & -0.0x0018 &  0.0x0004 & \pm~0.0x0001 &    -151.x06206\,(2) \\
    &               & $2p_{1/2}$ &   -148.x48661       &   0.1x0047 &  -0.0x0036 & -0.0x0008 & -0.0x0008 & \pm~0.0x0002 &    -148.x38666\,(2) \\
    &               & $2p_{3/2}$ &   -146.x04206       &   0.0x4201 &  -0.0x0539 &  0.0x0022 &  0.0x0003 & \pm~0.0x0002 &    -146.x00519\,(2) \\
 37 & 4.20x3\,(2)   & $2s_{1/2}$ &   -160.x17714       &   0.0x1110 &  -0.0x0734 & -0.0x0020 &  0.0x0005 & \pm~0.0x0001 &    -160.x17353\,(2) \\
    &               & $2p_{1/2}$ &   -157.x51436       &   0.1x1067 &  -0.0x0032 & -0.0x0009 & -0.0x0008 & \pm~0.0x0002 &    -157.x40418\,(2) \\
    &               & $2p_{3/2}$ &   -154.x76641       &   0.0x4630 &  -0.0x0554 &  0.0x0026 &  0.0x0003 & \pm~0.0x0002 &    -154.x72536\,(2) \\
 38 & 4.22x0\,(1)   & $2s_{1/2}$ &   -169.x57690       &   0.0x1450 &  -0.0x0748 & -0.0x0021 &  0.0x0005 & \pm~0.0x0001 &    -169.x57004\,(2) \\
    &               & $2p_{1/2}$ &   -166.x82632       &   0.1x2149 &  -0.0x0029 & -0.0x0009 & -0.0x0008 & \pm~0.0x0002 &    -166.x70529\,(2) \\
    &               & $2p_{3/2}$ &   -163.x74693       &   0.0x5083 &  -0.0x0569 &  0.0x0029 &  0.0x0003 & \pm~0.0x0002 &    -163.x70147\,(3) \\
 39 & 4.24x2\,(2)   & $2s_{1/2}$ &   -179.x26414       &   0.0x1817 &  -0.0x0762 & -0.0x0023 &  0.0x0005 & \pm~0.0x0002 &    -179.x25377\,(2) \\
    &               & $2p_{1/2}$ &   -176.x42466       &   0.1x3299 &  -0.0x0026 & -0.0x0010 & -0.0x0008 & \pm~0.0x0002 &    -176.x29211\,(2) \\
    &               & $2p_{3/2}$ &   -172.x98399       &   0.0x5559 &  -0.0x0584 &  0.0x0033 &  0.0x0003 & \pm~0.0x0003 &    -172.x93388\,(3) \\
 40 & 4.27x0\,(1)   & $2s_{1/2}$ &   -189.x24117       &   0.0x2212 &  -0.0x0777 & -0.0x0025 &  0.0x0005 & \pm~0.0x0002 &    -189.x22702\,(2) \\
    &               & $2p_{1/2}$ &   -186.x31165       &   0.1x4518 &  -0.0x0024 & -0.0x0010 & -0.0x0009 & \pm~0.0x0002 &    -186.x16690\,(2) \\
    &               & $2p_{3/2}$ &   -182.x47797       &   0.0x6060 &  -0.0x0600 &  0.0x0037 &  0.0x0003 & \pm~0.0x0003 &    -182.x42297\,(3) \\
 41 & 4.32x4\,(2)   & $2s_{1/2}$ &   -199.x51035       &   0.0x2636 &  -0.0x0792 & -0.0x0026 &  0.0x0005 & \pm~0.0x0002 &    -199.x49212\,(2) \\
    &               & $2p_{1/2}$ &   -196.x48962       &   0.1x5808 &  -0.0x0023 & -0.0x0010 & -0.0x0009 & \pm~0.0x0002 &    -196.x33196\,(3) \\
    &               & $2p_{3/2}$ &   -192.x22925       &   0.0x6585 &  -0.0x0617 &  0.0x0042 &  0.0x0003 & \pm~0.0x0003 &    -192.x16913\,(3) \\
 42 & 4.40x9\,(1)   & $2s_{1/2}$ &   -210.x07409       &   0.0x3089 &  -0.0x0808 & -0.0x0028 &  0.0x0005 & \pm~0.0x0002 &    -210.x05151\,(2) \\
    &               & $2p_{1/2}$ &   -206.x96097       &   0.1x7172 &  -0.0x0022 & -0.0x0011 & -0.0x0009 & \pm~0.0x0003 &    -206.x78967\,(3) \\
    &               & $2p_{3/2}$ &   -202.x23825       &   0.0x7135 &  -0.0x0634 &  0.0x0047 &  0.0x0003 & \pm~0.0x0003 &    -202.x17274\,(4) \\
 43 & 4.42x4\,(44)  & $2s_{1/2}$ &   -220.x93510\,(8)  &   0.0x3569 &  -0.0x0823 & -0.0x0030 &  0.0x0005 & \pm~0.0x0002 &    -220.x90789\,(8)  \\
    &               & $2p_{1/2}$ &   -217.x72825       &   0.1x8610 &  -0.0x0022 & -0.0x0011 & -0.0x0009 & \pm~0.0x0003 &    -217.x54257\,(3) \\
    &               & $2p_{3/2}$ &   -212.x50536       &   0.0x7711 &  -0.0x0652 &  0.0x0053 &  0.0x0003 & \pm~0.0x0004 &    -212.x43422\,(4) \\
 44 & 4.48x2\,(2)   & $2s_{1/2}$ &   -232.x09593       &   0.0x4086 &  -0.0x0840 & -0.0x0032 &  0.0x0006 & \pm~0.0x0002 &    -232.x06372\,(2) \\
    &               & $2p_{1/2}$ &   -228.x79408       &   0.2x0132 &  -0.0x0022 & -0.0x0011 & -0.0x0009 & \pm~0.0x0003 &    -228.x59319\,(3) \\
    &               & $2p_{3/2}$ &   -223.x03102       &   0.0x8313 &  -0.0x0671 &  0.0x0059 &  0.0x0002 & \pm~0.0x0004 &    -222.x95399\,(4) \\
 45 & 4.49x4\,(2)   & $2s_{1/2}$ &   -243.x55937       &   0.0x4627 &  -0.0x0856 & -0.0x0034 &  0.0x0006 & \pm~0.0x0002 &    -243.x52194\,(3) \\
    &               & $2p_{1/2}$ &   -240.x16111       &   0.2x1728 &  -0.0x0023 & -0.0x0011 & -0.0x0010 & \pm~0.0x0003 &    -239.x94427\,(3) \\
    &               & $2p_{3/2}$ &   -233.x81565       &   0.0x8940 &  -0.0x0690 &  0.0x0066 &  0.0x0002 & \pm~0.0x0004 &    -233.x73247\,(5) \\
 46 & 4.53x2\,(3)   & $2s_{1/2}$ &   -255.x32827       &   0.0x5205 &  -0.0x0873 & -0.0x0036 &  0.0x0006 & \pm~0.0x0002 &    -255.x28526\,(3) \\
    &               & $2p_{1/2}$ &   -251.x83223       &   0.2x3411 &  -0.0x0025 & -0.0x0011 & -0.0x0010 & \pm~0.0x0003 &    -251.x59857\,(4) \\
    &               & $2p_{3/2}$ &   -244.x85971       &   0.0x9595 &  -0.0x0710 &  0.0x0073 &  0.0x0002 & \pm~0.0x0005 &    -244.x77012\,(5) \\
 47 & 4.54x4\,(4)   & $2s_{1/2}$ &   -267.x40564\,(1)  &   0.0x5814 &  -0.0x0891 & -0.0x0038 &  0.0x0006 & \pm~0.0x0003 &    -267.x35673\,(3)  \\
    &               & $2p_{1/2}$ &   -263.x81030       &   0.2x5179 &  -0.0x0027 & -0.0x0011 & -0.0x0010 & \pm~0.0x0004 &    -263.x55899\,(4) \\
    &               & $2p_{3/2}$ &   -256.x16366       &   0.1x0276 &  -0.0x0731 &  0.0x0080 &  0.0x0002 & \pm~0.0x0005 &    -256.x06738\,(5) \\
 48 & 4.61x4\,(2)   & $2s_{1/2}$ &   -279.x79438\,(1)  &   0.0x6459 &  -0.0x0909 & -0.0x0040 &  0.0x0006 & \pm~0.0x0003 &    -279.x73921\,(3)  \\
    &               & $2p_{1/2}$ &   -276.x09840       &   0.2x7037 &  -0.0x0030 & -0.0x0011 & -0.0x0010 & \pm~0.0x0004 &    -275.x82853\,(4) \\
    &               & $2p_{3/2}$ &   -267.x72796       &   0.1x0984 &  -0.0x0752 &  0.0x0089 &  0.0x0002 & \pm~0.0x0006 &    -267.x62473\,(6) \\
 49 & 4.61x7\,(2)   & $2s_{1/2}$ &   -292.x49801\,(1)  &   0.0x7139 &  -0.0x0927 & -0.0x0042 &  0.0x0006 & \pm~0.0x0003 &    -292.x43624\,(3)  \\
    &               & $2p_{1/2}$ &   -288.x69965       &   0.2x8988 &  -0.0x0033 & -0.0x0011 & -0.0x0010 & \pm~0.0x0004 &    -288.x41032\,(4) \\
    &               & $2p_{3/2}$ &   -279.x55310       &   0.1x1720 &  -0.0x0774 &  0.0x0097 &  0.0x0002 & \pm~0.0x0006 &    -279.x44265\,(6) \\
 50 & 4.65x4\,(1)   & $2s_{1/2}$ &   -305.x51949\,(1)  &   0.0x7853 &  -0.0x0946 & -0.0x0045 &  0.0x0007 & \pm~0.0x0003 &    -305.x45080\,(3)  \\
    &               & $2p_{1/2}$ &   -301.x61727       &   0.3x1030 &  -0.0x0038 & -0.0x0010 & -0.0x0011 & \pm~0.0x0005 &    -301.x30756\,(5) \\
    &               & $2p_{3/2}$ &   -291.x63957       &   0.1x2483 &  -0.0x0797 &  0.0x0107 &  0.0x0002 & \pm~0.0x0006 &    -291.x52162\,(7) \\
 51 & 4.68x0\,(2)   & $2s_{1/2}$ &   -318.x86248\,(2)  &   0.0x8610 &  -0.0x0965 & -0.0x0047 &  0.0x0007 & \pm~0.0x0003 &    -318.x78643\,(4)  \\
    &               & $2p_{1/2}$ &   -314.x85471       &   0.3x3175 &  -0.0x0043 & -0.0x0010 & -0.0x0011 & \pm~0.0x0005 &    -314.x52359\,(5) \\
    &               & $2p_{3/2}$ &   -303.x98789       &   0.1x3275 &  -0.0x0821 &  0.0x0117 &  0.0x0002 & \pm~0.0x0007 &    -303.x86216\,(7) \\
 52 & 4.74x3\,(3)   & $2s_{1/2}$ &   -332.x53029\,(2)  &   0.0x9404 &  -0.0x0985 & -0.0x0049 &  0.0x0007 & \pm~0.0x0003 &    -332.x44652\,(4)  \\
    &               & $2p_{1/2}$ &   -328.x41543       &   0.3x5420 &  -0.0x0048 & -0.0x0010 & -0.0x0011 & \pm~0.0x0005 &    -328.x06192\,(5) \\
    &               & $2p_{3/2}$ &   -316.x59857       &   0.1x4094 &  -0.0x0845 &  0.0x0128 &  0.0x0001 & \pm~0.0x0007 &    -316.x46479\,(7) \\
 53 & 4.75x0\,(4)   & $2s_{1/2}$ &   -346.x52697\,(3)  &   0.1x0232 &  -0.0x1005 & -0.0x0051 &  0.0x0007 & \pm~0.0x0003 &    -346.x43514\,(5)  \\
    &               & $2p_{1/2}$ &   -342.x30302       &   0.3x7764 &  -0.0x0055 & -0.0x0009 & -0.0x0011 & \pm~0.0x0005 &    -341.x92614\,(6) \\
    &               & $2p_{3/2}$ &   -329.x47213       &   0.1x4942 &  -0.0x0871 &  0.0x0139 &  0.0x0001 & \pm~0.0x0008 &    -329.x33002\,(8) \\
 54 & 4.78x7\,(5)   & $2s_{1/2}$ &   -360.x85598\,(4)  &   0.1x1103 &  -0.0x1026 & -0.0x0054 &  0.0x0007 & \pm~0.0x0004 &    -360.x75567\,(5)  \\
    &               & $2p_{1/2}$ &   -356.x52130       &   0.4x0217 &  -0.0x0062 & -0.0x0009 & -0.0x0012 & \pm~0.0x0006 &    -356.x11995\,(6) \\
    &               & $2p_{3/2}$ &   -342.x60914       &   0.1x5818 &  -0.0x0897 &  0.0x0152 &  0.0x0001 & \pm~0.0x0008 &    -342.x45840\,(8) \\
 55 & 4.80x4\,(5)   & $2s_{1/2}$ &   -375.x52155\,(4)  &   0.1x2016 &  -0.0x1047 & -0.0x0056 &  0.0x0008 & \pm~0.0x0004 &    -375.x41234\,(6)  \\
    &               & $2p_{1/2}$ &   -371.x07417       &   0.4x2782 &  -0.0x0070 & -0.0x0008 & -0.0x0012 & \pm~0.0x0006 &    -370.x64725\,(6) \\
    &               & $2p_{3/2}$ &   -356.x01014       &   0.1x6723 &  -0.0x0924 &  0.0x0165 &  0.0x0001 & \pm~0.0x0009 &    -355.x85050\,(9) \\
 56 & 4.83x8\,(5)   & $2s_{1/2}$ &   -390.x52754\,(5)  &   0.1x2971 &  -0.0x1069 & -0.0x0058 &  0.0x0008 & \pm~0.0x0004 &    -390.x40902\,(6)  \\
    &               & $2p_{1/2}$ &   -385.x96566       &   0.4x5460 &  -0.0x0079 & -0.0x0008 & -0.0x0012 & \pm~0.0x0007 &    -385.x51205\,(7) \\
    &               & $2p_{3/2}$ &   -369.x67571       &   0.1x7656 &  -0.0x0952 &  0.0x0179 &  0.0x0000 & \pm~0.0x0009 &    -369.x50688\,(10) \\
 57 & 4.85x5\,(5)   & $2s_{1/2}$ &   -405.x87846\,(6)  &   0.1x3972 &  -0.0x1091 & -0.0x0061 &  0.0x0008 & \pm~0.0x0004 &    -405.x75018\,(7)  \\
    &               & $2p_{1/2}$ &   -401.x19999       &   0.4x8257 &  -0.0x0088 & -0.0x0007 & -0.0x0013 & \pm~0.0x0007 &    -400.x71850\,(7) \\
    &               & $2p_{3/2}$ &   -383.x60643       &   0.1x8618 &  -0.0x0981 &  0.0x0194 &  0.0x0000 & \pm~0.0x0010 &    -383.x42812\,(10) \\
 58 & 4.87x7\,(2)   & $2s_{1/2}$ &   -421.x57854\,(4)  &   0.1x5015 &  -0.0x1114 & -0.0x0063 &  0.0x0008 & \pm~0.0x0004 &    -421.x44008\,(6)  \\
    &               & $2p_{1/2}$ &   -416.x78149       &   0.5x1173 &  -0.0x0099 & -0.0x0006 & -0.0x0013 & \pm~0.0x0007 &    -416.x27094\,(7) \\
    &               & $2p_{3/2}$ &   -397.x80290       &   0.1x9609 &  -0.0x1011 &  0.0x0209 &  0.0x0000 & \pm~0.0x0011 &    -397.x61482\,(11) \\
 59 & 4.89x2\,(5)   & $2s_{1/2}$ &   -437.x63243\,(7)  &   0.1x6103 &  -0.0x1138 & -0.0x0065 &  0.0x0008 & \pm~0.0x0005 &    -437.x48335\,(9)  \\
    &               & $2p_{1/2}$ &   -432.x71467       &   0.5x4213 &  -0.0x0110 & -0.0x0005 & -0.0x0013 & \pm~0.0x0008 &    -432.x17382\,(8) \\
    &               & $2p_{3/2}$ &   -412.x26571       &   0.2x0629 &  -0.0x1042 &  0.0x0226 &  0.0x0000 & \pm~0.0x0011 &    -412.x06758\,(11) \\
 60 & 4.91x2\,(2)   & $2s_{1/2}$ &   -454.x04479\,(6)  &   0.1x7240 &  -0.0x1162 & -0.0x0068 &  0.0x0009 & \pm~0.0x0005 &    -453.x88459\,(8)  \\
    &               & $2p_{1/2}$ &   -449.x00422       &   0.5x7384 &  -0.0x0122 & -0.0x0004 & -0.0x0014 & \pm~0.0x0008 &    -448.x43178\,(8) \\
    &               & $2p_{3/2}$ &   -426.x99551       &   0.2x1678 &  -0.0x1073 &  0.0x0244 & -0.0x0001 & \pm~0.0x0012 &    -426.x78703\,(12) \\
 61 & 4.96x2\,(50)  & $2s_{1/2}$ &   -470.x82019\,(69) &   0.1x8416 &  -0.0x1187 & -0.0x0070 &  0.0x0009 & \pm~0.0x0005 &    -470.x64850\,(70) \\
    &               & $2p_{1/2}$ &   -465.x65499\,(3)  &   0.6x0677 &  -0.0x0136 & -0.0x0003 & -0.0x0014 & \pm~0.0x0009 &    -465.x04975\,(9)  \\
    &               & $2p_{3/2}$ &   -441.x99292       &   0.2x2756 &  -0.0x1106 &  0.0x0262 & -0.0x0001 & \pm~0.0x0013 &    -441.x77382\,(13) \\
 62 & 5.08x4\,(6)   & $2s_{1/2}$ &   -487.x96303\,(11) &   0.1x9662 &  -0.0x1212 & -0.0x0072 &  0.0x0009 & \pm~0.0x0005 &    -487.x77916\,(12) \\
    &               & $2p_{1/2}$ &   -482.x67204       &   0.6x4125 &  -0.0x0150 & -0.0x0002 & -0.0x0014 & \pm~0.0x0009 &    -482.x03246\,(9) \\
    &               & $2p_{3/2}$ &   -457.x25862       &   0.2x3862 &  -0.0x1141 &  0.0x0282 & -0.0x0001 & \pm~0.0x0013 &    -457.x02860\,(13) \\
 63 & 5.11x3\,(4)   & $2s_{1/2}$ &   -505.x48081\,(10) &   0.2x0935 &  -0.0x1238 & -0.0x0075 &  0.0x0009 & \pm~0.0x0005 &    -505.x28450\,(11) \\
    &               & $2p_{1/2}$ &   -500.x06064       &   0.6x7693 &  -0.0x0164 & -0.0x0001 & -0.0x0014 & \pm~0.0x0010 &    -499.x38550\,(10) \\
    &               & $2p_{3/2}$ &   -472.x79322       &   0.2x4997 &  -0.0x1176 &  0.0x0303 & -0.0x0002 & \pm~0.0x0014 &    -472.x55200\,(14) \\
 64 & 5.16x2\,(2)   & $2s_{1/2}$ &   -523.x37749\,(9)  &   0.2x2275 &  -0.0x1265 & -0.0x0077 &  0.0x0009 & \pm~0.0x0006 &    -523.x16807\,(11) \\
    &               & $2p_{1/2}$ &   -517.x82618       &   0.7x1421 &  -0.0x0180 &  0.0x0000 & -0.0x0015 & \pm~0.0x0010 &    -517.x11392\,(10) \\
    &               & $2p_{3/2}$ &   -488.x59744       &   0.2x6161 &  -0.0x1212 &  0.0x0325 & -0.0x0002 & \pm~0.0x0015 &    -488.x34473\,(15) \\
 65 & 5.06x0\,(150) & $2s_{1/2}$ &   -541.x6620\,(32)  &   0.2x3657 &  -0.0x1293 & -0.0x0079 &  0.0x0010 & \pm~0.0x0006 &    -541.x4391\,(32)  \\
    &               & $2p_{1/2}$ &   -535.x97441\,(16) &   0.7x5293 &  -0.0x0197 &  0.0x0001 & -0.0x0015 & \pm~0.0x0011 &    -535.x22358\,(19) \\
    &               & $2p_{3/2}$ &   -504.x67192       &   0.2x7353 &  -0.0x1249 &  0.0x0348 & -0.0x0002 & \pm~0.0x0015 &    -504.x40743\,(15) \\
 66 & 5.22x1\,(2)   & $2s_{1/2}$ &   -560.x33217\,(13) &   0.2x5100 &  -0.0x1321 & -0.0x0081 &  0.0x0010 & \pm~0.0x0006 &    -560.x09510\,(14) \\
    &               & $2p_{1/2}$ &   -554.x51094       &   0.7x9318 &  -0.0x0215 &  0.0x0003 & -0.0x0016 & \pm~0.0x0012 &    -553.x72003\,(12) \\
    &               & $2p_{3/2}$ &   -521.x01747       &   0.2x8573 &  -0.0x1288 &  0.0x0372 & -0.0x0003 & \pm~0.0x0016 &    -520.x74092\,(16) \\
 67 & 5.20x2\,(31)  & $2s_{1/2}$ &   -579.x40308\,(83) &   0.2x6596 &  -0.0x1350 & -0.0x0083 &  0.0x0010 & \pm~0.0x0006 &    -579.x15136\,(83) \\
    &               & $2p_{1/2}$ &   -573.x44216\,(4)  &   0.8x3506 &  -0.0x0234 &  0.0x0004 & -0.0x0016 & \pm~0.0x0012 &    -572.x60956\,(13)  \\
    &               & $2p_{3/2}$ &   -537.x63469       &   0.2x9821 &  -0.0x1328 &  0.0x0398 & -0.0x0003 & \pm~0.0x0017 &    -537.x34581\,(17) \\
 68 & 5.25x0\,(3)   & $2s_{1/2}$ &   -598.x87578\,(17) &   0.2x8145 &  -0.0x1380 & -0.0x0085 &  0.0x0010 & \pm~0.0x0007 &    -598.x60888\,(18) \\
    &               & $2p_{1/2}$ &   -592.x77424       &   0.8x7853 &  -0.0x0254 &  0.0x0006 & -0.0x0016 & \pm~0.0x0013 &    -591.x89835\,(13) \\
    &               & $2p_{3/2}$ &   -554.x52439       &   0.3x1097 &  -0.0x1369 &  0.0x0425 & -0.0x0004 & \pm~0.0x0018 &    -554.x22290\,(18) \\
 69 & 5.22x6\,(4)   & $2s_{1/2}$ &   -618.x76050\,(20) &   0.2x9754 &  -0.0x1411 & -0.0x0087 &  0.0x0011 & \pm~0.0x0007 &    -618.x47784\,(21) \\
    &               & $2p_{1/2}$ &   -612.x51384\,(1)  &   0.9x2373 &  -0.0x0274 &  0.0x0007 & -0.0x0017 & \pm~0.0x0014 &    -611.x59295\,(14)  \\
    &               & $2p_{3/2}$ &   -571.x68728       &   0.3x2401 &  -0.0x1412 &  0.0x0453 & -0.0x0004 & \pm~0.0x0019 &    -571.x37290\,(19) \\
 70 & 5.31x1\,(6)   & $2s_{1/2}$ &   -639.x05870\,(28) &   0.3x1424 &  -0.0x1443 & -0.0x0089 &  0.0x0011 & \pm~0.0x0007 &    -638.x75966\,(29) \\
    &               & $2p_{1/2}$ &   -632.x66766\,(2)  &   0.9x7068 &  -0.0x0296 &  0.0x0009 & -0.0x0017 & \pm~0.0x0014 &    -631.x70003\,(14)  \\
    &               & $2p_{3/2}$ &   -589.x12417       &   0.3x3731 &  -0.0x1456 &  0.0x0482 & -0.0x0005 & \pm~0.0x0019 &    -588.x79664\,(19) \\
 71 & 5.37x0\,(30)  & $2s_{1/2}$ &   -659.x7814\,(12)  &   0.3x3151 &  -0.0x1476 & -0.0x0090 &  0.0x0011 & \pm~0.0x0008 &    -659.x4655\,(12)  \\
    &               & $2p_{1/2}$ &   -653.x24294\,(7)  &   1.0x1940 &  -0.0x0319 &  0.0x0011 & -0.0x0017 & \pm~0.0x0015 &    -652.x22680\,(17)  \\
    &               & $2p_{3/2}$ &   -606.x83581       &   0.3x5088 &  -0.0x1501 &  0.0x0514 & -0.0x0005 & \pm~0.0x0020 &    -606.x49485\,(20) \\
 72 & 5.34x2\,(2)   & $2s_{1/2}$ &   -680.x93893\,(26) &   0.3x4931 &  -0.0x1509 & -0.0x0092 &  0.0x0011 & \pm~0.0x0008 &    -680.x60551\,(28) \\
    &               & $2p_{1/2}$ &   -674.x24721\,(2)  &   1.0x6994 &  -0.0x0344 &  0.0x0013 & -0.0x0018 & \pm~0.0x0016 &    -673.x18076\,(16)  \\
    &               & $2p_{3/2}$ &   -624.x82296       &   0.3x6472 &  -0.0x1547 &  0.0x0546 & -0.0x0006 & \pm~0.0x0021 &    -624.x46831\,(21) \\
 73 & 5.35x1\,(3)   & $2s_{1/2}$ &   -702.x53427\,(32) &   0.3x6780 &  -0.0x1544 & -0.0x0093 &  0.0x0012 & \pm~0.0x0008 &    -702.x18272\,(33) \\
    &               & $2p_{1/2}$ &   -695.x68798\,(2)  &   1.1x2246 &  -0.0x0369 &  0.0x0014 & -0.0x0018 & \pm~0.0x0017 &    -694.x56925\,(17)  \\
    &               & $2p_{3/2}$ &   -643.x08648       &   0.3x7881 &  -0.0x1595 &  0.0x0580 & -0.0x0006 & \pm~0.0x0022 &    -642.x71788\,(22) \\
 74 & 5.36x7\,(2)   & $2s_{1/2}$ &   -724.x57660\,(32) &   0.3x8692 &  -0.0x1579 & -0.0x0094 &  0.0x0012 & \pm~0.0x0008 &    -724.x20630\,(33) \\
    &               & $2p_{1/2}$ &   -717.x57343\,(2)  &   1.1x7695 &  -0.0x0395 &  0.0x0016 & -0.0x0019 & \pm~0.0x0017 &    -716.x40046\,(18)  \\
    &               & $2p_{3/2}$ &   -661.x62718       &   0.3x9316 &  -0.0x1645 &  0.0x0616 & -0.0x0007 & \pm~0.0x0023 &    -661.x24438\,(23) \\
 75 & 5.33x9\,(13)  & $2s_{1/2}$ &   -747.x07707\,(85) &   0.4x0664 &  -0.0x1616 & -0.0x0095 &  0.0x0012 & \pm~0.0x0009 &    -746.x68742\,(85) \\
    &               & $2p_{1/2}$ &   -739.x91201\,(6)  &   1.2x3347 &  -0.0x0423 &  0.0x0018 & -0.0x0019 & \pm~0.0x0018 &    -738.x68278\,(19)  \\
    &               & $2p_{3/2}$ &   -680.x44589       &   0.4x0775 &  -0.0x1696 &  0.0x0653 & -0.0x0007 & \pm~0.0x0024 &    -680.x04863\,(24) \\
 76 & 5.41x3\,(1)   & $2s_{1/2}$ &   -770.x03598\,(40) &   0.4x2704 &  -0.0x1654 & -0.0x0096 &  0.0x0013 & \pm~0.0x0009 &    -769.x62631\,(41) \\
    &               & $2p_{1/2}$ &   -762.x71201\,(3)  &   1.2x9208 &  -0.0x0452 &  0.0x0020 & -0.0x0020 & \pm~0.0x0019 &    -761.x42444\,(19)  \\
    &               & $2p_{3/2}$ &   -699.x54352       &   0.4x2259 &  -0.0x1748 &  0.0x0692 & -0.0x0008 & \pm~0.0x0025 &    -699.x13157\,(25) \\
 77 & 5.40x2\,(106) & $2s_{1/2}$ &   -793.x4734\,(80)  &   0.4x4808 &  -0.0x1693 & -0.0x0097 &  0.0x0013 & \pm~0.0x0009 &    -793.x0431\,(80)  \\
    &               & $2p_{1/2}$ &   -785.x98315\,(59) &   1.3x5290 &  -0.0x0482 &  0.0x0022 & -0.0x0020 & \pm~0.0x0020 &    -784.x63506\,(62) \\
    &               & $2p_{3/2}$ &   -718.x92087       &   0.4x3767 &  -0.0x1802 &  0.0x0733 & -0.0x0009 & \pm~0.0x0026 &    -718.x49398\,(26) \\
 78 & 5.42x8\,(3)   & $2s_{1/2}$ &   -817.x39139\,(55) &   0.4x6992 &  -0.0x1733 & -0.0x0097 &  0.0x0013 & \pm~0.0x0010 &    -816.x93964\,(56) \\
    &               & $2p_{1/2}$ &   -809.x73458\,(4)  &   1.4x1607 &  -0.0x0513 &  0.0x0024 & -0.0x0021 & \pm~0.0x0021 &    -808.x32362\,(21)  \\
    &               & $2p_{3/2}$ &   -738.x57887       &   0.4x5297 &  -0.0x1858 &  0.0x0775 & -0.0x0009 & \pm~0.0x0027 &    -738.x13682\,(27) \\
 79 & 5.43x6\,(4)   & $2s_{1/2}$ &   -841.x80344\,(67) &   0.4x9238 &  -0.0x1774 & -0.0x0097 &  0.0x0014 & \pm~0.0x0010 &    -841.x32964\,(67) \\
    &               & $2p_{1/2}$ &   -833.x97640\,(5)  &   1.4x8152 &  -0.0x0546 &  0.0x0027 & -0.0x0021 & \pm~0.0x0022 &    -832.x50029\,(23)  \\
    &               & $2p_{3/2}$ &   -758.x51842       &   0.4x6850 &  -0.0x1916 &  0.0x0820 & -0.0x0010 & \pm~0.0x0028 &    -758.x06098\,(28) \\
 80 & 5.46x3\,(2)   & $2s_{1/2}$ &   -866.x71677\,(67) &   0.5x1557 &  -0.0x1817 & -0.0x0097 &  0.0x0014 & \pm~0.0x0010 &    -866.x22020\,(67) \\
    &               & $2p_{1/2}$ &   -858.x71879\,(5)  &   1.5x4942 &  -0.0x0580 &  0.0x0029 & -0.0x0022 & \pm~0.0x0023 &    -857.x17510\,(24)  \\
    &               & $2p_{3/2}$ &   -778.x74045       &   0.4x8423 &  -0.0x1975 &  0.0x0866 & -0.0x0011 & \pm~0.0x0029 &    -778.x26741\,(29) \\
 81 & 5.47x6\,(3)   & $2s_{1/2}$ &   -892.x14506\,(79) &   0.5x3945 &  -0.0x1861 & -0.0x0096 &  0.0x0014 & \pm~0.0x0011 &    -891.x62504\,(79) \\
    &               & $2p_{1/2}$ &   -883.x97272\,(6)  &   1.6x1981 &  -0.0x0615 &  0.0x0031 & -0.0x0023 & \pm~0.0x0024 &    -882.x35898\,(25)  \\
    &               & $2p_{3/2}$ &   -799.x24589       &   0.5x0018 &  -0.0x2035 &  0.0x0915 & -0.0x0011 & \pm~0.0x0030 &    -798.x75703\,(30) \\
 82 & 5.50x1\,(1)   & $2s_{1/2}$ &   -918.x09688\,(83) &   0.5x6397 &  -0.0x1906 & -0.0x0096 &  0.0x0015 & \pm~0.0x0011 &    -917.x55279\,(84) \\
    &               & $2p_{1/2}$ &   -909.x74938\,(7)  &   1.6x9271 &  -0.0x0652 &  0.0x0033 & -0.0x0023 & \pm~0.0x0025 &    -908.x06309\,(26)  \\
    &               & $2p_{3/2}$ &   -820.x03570       &   0.5x1632 &  -0.0x2098 &  0.0x0965 & -0.0x0012 & \pm~0.0x0031 &    -819.x53083\,(31) \\
 83 & 5.52x1\,(3)   & $2s_{1/2}$ &   -944.x5857\,(10)  &   0.5x8922 &  -0.0x1953 & -0.0x0095 &  0.0x0015 & \pm~0.0x0012 &    -944.x0168\,(10)  \\
    &               & $2p_{1/2}$ &   -936.x06078\,(9)  &   1.7x6830 &  -0.0x0690 &  0.0x0035 & -0.0x0024 & \pm~0.0x0026 &    -934.x29927\,(28) \\
    &               & $2p_{3/2}$ &   -841.x11086       &   0.5x3265 &  -0.0x2162 &  0.0x1017 & -0.0x0013 & \pm~0.0x0032 &    -840.x58979\,(32) \\
 84 & 5.52x6\,(13)  & $2s_{1/2}$ &   -971.x6251\,(22)  &   0.6x1524 &  -0.0x2002 & -0.0x0093 &  0.0x0015 & \pm~0.0x0012 &    -971.x0307\,(22)  \\
    &               & $2p_{1/2}$ &   -962.x91947\,(20) &   1.8x4671 &  -0.0x0730 &  0.0x0037 & -0.0x0024 & \pm~0.0x0028 &    -961.x07993\,(34) \\
    &               & $2p_{3/2}$ &   -862.x47235       &   0.5x4915 &  -0.0x2229 &  0.0x1072 & -0.0x0014 & \pm~0.0x0034 &    -861.x93490\,(34) \\
 85 & 5.53x9\,(55)  & $2s_{1/2}$ &   -999.x2252\,(94)  &   0.6x4207 &  -0.0x2052 & -0.0x0092 &  0.0x0016 & \pm~0.0x0012 &    -998.x6044\,(94)  \\
    &               & $2p_{1/2}$ &   -990.x33823\,(89) &   1.9x2808 &  -0.0x0771 &  0.0x0039 & -0.0x0025 & \pm~0.0x0029 &    -988.x41772\,(93) \\
    &               & $2p_{3/2}$ &   -884.x12119       &   0.5x6583 &  -0.0x2297 &  0.0x1129 & -0.0x0014 & \pm~0.0x0035 &    -883.x56719\,(35) \\
 86 & 5.65x5\,(16)  & $2s_{1/2}$ &  -1027.x3822\,(33)  &   0.6x6963 &  -0.0x2104 & -0.0x0090 &  0.0x0016 & \pm~0.0x0013 &   -1026.x7343\,(33)  \\
    &               & $2p_{1/2}$ &  -1018.x32939\,(33) &   2.0x1237 &  -0.0x0814 &  0.0x0041 & -0.0x0026 & \pm~0.0x0030 &   -1016.x32500\,(44) \\
    &               & $2p_{3/2}$ &   -906.x05854       &   0.5x8266 &  -0.0x2367 &  0.0x1188 & -0.0x0015 & \pm~0.0x0036 &    -905.x48783\,(36) \\
 87 & 5.65x8\,(13)  & $2s_{1/2}$ &  -1056.x1458\,(31)  &   0.6x9798 &  -0.0x2157 & -0.0x0087 &  0.0x0017 & \pm~0.0x0013 &   -1055.x4701\,(31)  \\
    &               & $2p_{1/2}$ &  -1046.x91008\,(31) &   2.0x9987 &  -0.0x0858 &  0.0x0043 & -0.0x0026 & \pm~0.0x0031 &   -1044.x81862\,(44) \\
    &               & $2p_{3/2}$ &   -928.x28519       &   0.5x9963 &  -0.0x2439 &  0.0x1249 & -0.0x0016 & \pm~0.0x0037 &    -927.x69762\,(37) \\
 88 & 5.68x4\,(26)  & $2s_{1/2}$ &  -1085.x5082\,(62)  &   0.7x2710 &  -0.0x2212 & -0.0x0084 &  0.0x0017 & \pm~0.0x0014 &   -1084.x8039\,(62)  \\
    &               & $2p_{1/2}$ &  -1076.x09350\,(65) &   2.1x9058 &  -0.0x0904 &  0.0x0045 & -0.0x0027 & \pm~0.0x0033 &   -1073.x91178\,(73) \\
    &               & $2p_{3/2}$ &   -950.x80236       &   0.6x1674 &  -0.0x2513 &  0.0x1313 & -0.0x0017 & \pm~0.0x0038 &    -950.x19780\,(38) \\
 89 & 5.67x0\,(57)  & $2s_{1/2}$ &  -1115.x498\,(14)   &   0.7x5687 &  -0.0x2269 & -0.0x0081 &  0.0x0017 & \pm~0.0x0014 &   -1114.x764\,(14)   \\
    &               & $2p_{1/2}$ &  -1105.x8966\,(16)  &   2.2x8455 &  -0.0x0951 &  0.0x0047 & -0.0x0028 & \pm~0.0x0034 &   -1103.x6214\,(16)  \\
    &               & $2p_{3/2}$ &   -973.x61105       &   0.6x3395 &  -0.0x2590 &  0.0x1379 & -0.0x0018 & \pm~0.0x0039 &    -972.x98938\,(39) \\
 90 & 5.71x0\,(50)  & $2s_{1/2}$ &  -1146.x109\,(14)   &   0.7x8750 &  -0.0x2328 & -0.0x0077 &  0.0x0018 & \pm~0.0x0015 &   -1145.x346\,(14)   \\
    &               & $2p_{1/2}$ &  -1136.x3341\,(16)  &   2.3x8209 &  -0.0x1000 &  0.0x0049 & -0.0x0028 & \pm~0.0x0035 &   -1133.x9618\,(16)  \\
    &               & $2p_{3/2}$ &   -996.x71252       &   0.6x5128 &  -0.0x2668 &  0.0x1448 & -0.0x0019 & \pm~0.0x0041 &    -996.x07363\,(41) \\
 91 & 5.70x0\,(57)  & $2s_{1/2}$ &  -1177.x385\,(18)   &   0.8x1929 &  -0.0x2385 & -0.0x0072 &  0.0x0018 & \pm~0.0x0015 &   -1176.x590\,(18)   \\
    &               & $2p_{1/2}$ &  -1167.x4260\,(20)  &   2.4x8333 &  -0.0x1052 &  0.0x0051 & -0.0x0029 & \pm~0.0x0037 &   -1164.x9529\,(21)  \\
    &               & $2p_{3/2}$ &  -1020.x10774       &   0.6x6872 &  -0.0x2749 &  0.0x1519 & -0.0x0020 & \pm~0.0x0042 &   -1019.x45151\,(42) \\
 92 & 5.85x1\,(7)   & $2s_{1/2}$ &  -1209.x2997\,(37)  &   0.8x7286 &  -0.0x2466 & -0.0x0067 &  0.0x0019 & \pm~0.0x0016 &   -1208.x4520\,(37)  \\
    &               & $2p_{1/2}$ &  -1199.x18416\,(42) &   2.5x8825 &  -0.0x1104 &  0.0x0053 & -0.0x0030 & \pm~0.0x0038 &   -1196.x60672\,(57) \\
    &               & $2p_{3/2}$ &  -1043.x79820       &   0.6x8624 &  -0.0x2831 &  0.0x1593 & -0.0x0021 & \pm~0.0x0043 &   -1043.x12456\,(43) \\

\hline\hline
\end{longtable*}
\end{ruledtabular}
\endgroup

\end{document}